\documentclass[12pt]{article}
\usepackage[utf8]{inputenc}
\usepackage[letterpaper]{geometry}
\usepackage{color}
\usepackage{amsmath}
\usepackage{amssymb}
\usepackage{graphicx}
\usepackage{setspace}
\usepackage{esint}
\usepackage[authoryear,longnamesfirst]{natbib}
\usepackage[bookmarks=true,bookmarksnumbered=false,bookmarksopen=true,bookmarksopenlevel=1,
 breaklinks=true,pdfborder={0 0 1},backref=false,colorlinks=true]
 {hyperref}
\hypersetup{
 citecolor=blue}

\makeatletter

\providecommand{\tabularnewline}{\\}

\usepackage{amsfonts}\usepackage{mathrsfs}\usepackage{bm}\usepackage{verbatim}\usepackage{dsfont}\usepackage{tabularx}
\usepackage{lipsum}\usepackage{ltablex}\usepackage{array}\usepackage{makecell}\usepackage{caption}\usepackage{multirow}\usepackage{wasysym}\usepackage{soul}\usepackage{rotating}
\usepackage[hang,flushmargin]{footmisc}

\usepackage{algorithm}
\usepackage{algpseudocodex}
\normalbaselines 

\providecommand{\U}[1]{\protect\rule{.1in}{.1in}}
\newtheorem{theorem}{Theorem}[section]

\newtheorem{assumption}{Assumption}

\newtheorem{corollary}[theorem]{Corollary}

\newtheorem{lemma}[theorem]{Lemma}
\newtheorem{remark}{Remark}

\@addtoreset{lemma}{Appendix B}

\numberwithin{equation}{section}

\newcommand{\Keywords}[1]{\par\noindent {\small{\em Keywords\/}: #1}} 

\makeatother

\begin{document}
\title{On the Rank Condition of Tail Index Regressions and a Comparative
Study with Extremal Quantile Regression\thanks{This paper was previously circulated under the title \textquotedblleft Cautions
on Tail Index Regressions and a Comparative Study with Extremal Quantile
Regression\textquotedblright . We thank the editor, Viktor Todorov,
an associate editor, and two anonymous referees for their constructive
comments and valuable suggestions, which have led to a substantially
improved article. We are grateful to Yonghui Zhang for helpful discussions
and Jianhua Mei for research assistance. The replication package is
available at: https://github.com/youngtigers/Tail-index-regressions.}}
\author{{Thomas T. Yang}\thanks{Corresponding author. The Australian National University, Canberra,
ACT 0200, Australia. Email: \protect\protect\protect\protect\protect\protect\protect\protect\protect\protect\protect\protect\protect\protect\protect\protect\protect\protect\protect\protect\protect\protect\protect\protect\protect\protect\protect\protect\protect\protect\protect\protect\protect\protect\protect\protect\protect\href{http://tao.yang@anu.edu.au}{tao.yang@anu.edu.au}.} \\
 {\normalsize{}{}{}{}{}Australian National University }}
\date{Oct 2026}

\maketitle
\onehalfspacing 
\begin{abstract}
We revisit tail index regressions. For linear specifications, we find
that the usual full rank condition can fail because conditioning on
extreme outcomes causes regressors to degenerate to constants. Taking
this into account, we provide additional regularity conditions and
establish the corresponding asymptotic theory. For more general specifications,
the conditional distribution of the covariates in the tails concentrates
on the values that minimize the tail index. This issue does not arise
in the extremal quantile regression framework, where the tail index
is assumed constant. Simulations support these findings. Using daily
S\&P 500 returns, we give an empirical illustration. The patterns
we find are more consistent with a constant tail index and a time-varying
scale than with the strong degeneracy implied by a varying tail index.

\vspace{4mm}

\medskip{}

\vspace{2cm}

\end{abstract}

\setstretch{1.6}

\section{Introduction}

Tail index estimation has been widely applied in modeling extreme
events, such as natural environmental and financial market extremes;
see, for example, \citet{de_Haan_Ferreira} and references within.
Building on the seminal paper of \citet{Hill1975}, subsequent work
has sought to model the tail index as a function of covariates, either
parametrically (e.g., \citet{HallTajvidi} and \citet{WangTsai2009}),
semiparametrically (e.g., \citet{LiLengYou2022}), or nonparametrically
(e.g., \citet{deHaan03072021}). The literature is large; we mention
only a few important and representative papers. In this paper, we
study the rank condition and related conditions in tail index regressions.

To set the stage, suppose we observe $(Y,X)$, where $X$ enters the
tail index function of $Y$. Following the literature, we assume 
\begin{equation}
\bar{F}(y\mid x)=1-F(y\mid x)=y^{-\alpha(x)}L(y;x),\quad\text{for }y\geq w,\label{eq:surviveCDF}
\end{equation}
where $F(y\mid x)$ is the cumulative distribution function of $Y$
conditional on $X=x,$ and $L(y;x)$ is a slowly varying function
that satisfies $L(yt;x)/L(y;x)\to1$ for any $t>0$ as $y\to\infty$.
Adding $L(y;x)$ makes the distributional assumption in \eqref{eq:surviveCDF}
much more general than one without it; see \citet{de_Haan_Ferreira}
and \citet{WangTsai2009} for discussion.

Because $L(y;x)$ is present, maximum likelihood estimation based
on the approximation $\bar{F}(y\mid x)\approx Cy^{-\alpha(x)}\textrm{ for some }C>0$
has a bias term, which goes to zero only as $w\to\infty$. Therefore,
as the sample size grows, one must take a larger threshold $w$ and
restrict the analysis to samples with $Y>w$, so that the bias disappears
asymptotically.

We find that the full rank condition on $\mathbb{E}(XX'|Y>w)$ is
likely to fail asymptotically in tail index regressions as $w\rightarrow\infty$.
A related concentration problem arises in semiparametric tail index
regressions, where it can weaken identification of the nonparametric
component. Fully nonparametric methods based on local observations
are less affected because the tail index behaves approximately like
a constant within a shrinking neighborhood. The intuition follows
from Bayes’ theorem: conditional on large values of $Y$, the distribution
of $X$ degenerates toward the point $x$ where $\alpha(x)$ attains
its minimum, because extreme events are more likely to occur when
$\alpha(x)$ is smaller. Since only observations with $Y>w$ are retained
and the threshold $w\to\infty$ as the sample size grows, the variation
in $X$ shrinks, and the full rank condition fails. We note that the
main results in \citet{WangTsai2009} are correct because their asymptotics
can accommodate an asymptotically degenerate $\mathbb{E}(XX'|Y>w)$.
The purpose of this paper is to study this condition explicitly in
detail to avoid misunderstandings in future work in this line of research.

Specifically, the main results on tail index regressions are as follows: 
\begin{enumerate}
\item In the parametric case where $\alpha(X)=\exp\left(X'\theta^{*}\right)$,
with the $\exp$ function used to keep the index positive, we show
that $\mathbb{E}(XX'\mid Y>w)$ becomes nearly singular as $w$ grows
large, under fairly general conditions, even when $\mathbb{E}(XX')$
is nonsingular. As a consequence, the convergence rate of the parametric
estimator in \citet{WangTsai2009} is slower in the degenerate directions.
We provide conditions under which the nearly singular $\mathbb{E}(XX'\mid Y>w)$
can still be accommodated, and establish the corresponding asymptotic
properties. 
\item In semiparametric cases where $\alpha(X)$ is a general smooth function
of $X$, we show that the conditional density $f(x\mid Y>w)$ converges
to zero for all $x$ except at points where $\alpha(x)$ attains its
minimum, even though the unconditional density $f(x)$ is uniformly
bounded away from zero. This property may make it difficult to estimate
the nonparametric component in semiparametric models. 
\item The nonparametric setting is less affected by this issue, because
only local observations are used and $\alpha(\cdot)$ behaves like
a constant locally. 
\end{enumerate}
As a comparison, we investigate this rank condition for the extremal
quantile regression, introduced by \citet{Chernozhukov2005}. The
framework in \citet{Chernozhukov2005} assumed the conditional survival
function satisfies 
\[
\bar{F}\left(y|x\right)=k\left(x\right)\cdot y^{-\alpha}L\left(y\right),
\]
where $k\left(x\right)$ is some positive function, and $L\left(y\right)$
is a slowly varying function. The key difference from tail index regressions
is that $\alpha$ is a constant, so that all the heterogeneity comes
from $k\left(x\right)$. We find that the rank condition issue on
$\mathbb{E}(XX'|Y>w)$ does not show up for extremal quantile regression
for very general $k\left(x\right).$ Although the limiting conditional
distribution of $X|\left(Y>w\right)$ is tilted toward larger values
of $k\left(x\right)$, it does not degenerate. This shows the source
of the rank condition issue: a varying tail index $\alpha$.

Small-scale simulation studies confirm these findings. We further
study extreme daily losses (negative returns) of the S\&P 500 over
a long time horizon. As losses become more extreme, we find that such
events become more concentrated around major financial crises, but
the conditional variance of the timing of these extreme events does
not seem to degenerate. The empirical patterns are more consistent
with a constant tail index and a time-varying $k$, although the diagnostic
does not formally distinguish variation in the tail index from variation
in the scale function.

The rest of the paper is structured as follows. Section~\ref{sec:2}
formalizes the results for tail index regressions. Section \ref{sec:compare}
investigates the extremal quantile regression. Sections \ref{sec:Simulations}
and \ref{sec:Empirical-Illustration} provide small-sample evidence
from simulations and an empirical illustration, respectively. Section~\ref{sec:Conclusion}
concludes. Proofs and technical lemmas are collected in the Appendix.

\textit{Notation.} $\log\left(\cdot\right)$ stands for the natural
logarithm. For a vector $x$, $\left\Vert x\right\Vert $ denotes
the $L_{2}$ norm, and for a matrix $A,$ $\left\Vert A\right\Vert $
denotes the spectral norm. $\rho_{\min}(\cdot)$ and $\rho_{\max}(\cdot)$
denote the minimum and the maximum eigenvalues of a matrix, respectively.
$\mathbb{I}_{m}$ denotes the $m\times m$ identity matrix. For a
positive definite matrix $A$, $A^{1/2}$ denotes its symmetric positive-definite
square root. For the deterministic series $\{a_{n},b_{n}\}_{n=1}^{\infty}$,
we denote $a_{n}\propto b_{n}$ if $0<C_{1}\leq\liminf_{n\rightarrow\infty}\left\vert a_{n}/b_{n}\right\vert \leq\limsup_{n\rightarrow\infty}\left\vert a_{n}/b_{n}\right\vert \leq C_{2}<\infty$
for some constants $C_{1}$ and $C_{2}$ , $a_{n}\ll b_{n}$ if $a_{n}=o(b_{n})$,
and $a_{n}\gg b_{n}$ if $b_{n}\ll a_{n}$. $\overset{P}{\rightarrow}$
and $\overset{d}{\rightarrow}$ denote convergence in probability
and distribution, respectively. $C$ denotes some generic positive
constants that may vary from line to line.

\section{Models and the Rank Condition}

\label{sec:2}

We investigate the right tail behavior of $Y$ conditional on $X.$
To simplify the presentation, we assume that $Y$ is unbounded above
conditional on any $X$ and has a heavy-tailed distribution.

\subsection{The Linear Model}

For model (\ref{eq:surviveCDF}), \citet{WangTsai2009} assumed that
$\alpha(X)=\exp\left(X'\theta^{*}\right)$ and that the observations
$(x_{i},y_{i}),\,i=1,2,\ldots,n$, are i.i.d. For finite samples,
the truncation parameter is allowed to depend on $n$, denoted by
$w_{n}$. They proposed estimating $\theta^{*}$ by minimizing the
approximate negative log-likelihood function: 
\[
\hat{\theta}=\arg\min_{\theta}\sum_{i=1}^{n}\Bigl\{\exp(x_{i}'\theta)\,\log\!\left(\frac{y_{i}}{w_{n}}\right)-x_{i}'\theta\Bigr\}\,I(y_{i}>w_{n}),
\]
where $I(\cdot)$ is the indicator function.

A key condition in \citet{WangTsai2009} for identification of $\theta^{*}$
is that the Gram matrix 
\[
\hat{\varSigma}_{w_{n}}=\frac{1}{n_{0}}\sum_{i=1}^{n}x_{i}x_{i}'I(y_{i}>w_{n}),\quad\text{with }n_{0}=\sum_{i=1}^{n}I(y_{i}>w_{n}),
\]
is non-singular.

We will show that under fairly general conditions, the minimum eigenvalue
of the population counterpart 
\begin{equation}
\bar{\varSigma}_{w_{n}}=\mathbb{E}\left(\left.XX'\right|Y>w_{n}\right)\label{eq:Sigma_w_bar}
\end{equation}
converges to zero even if $\mathbb{E}(XX')$ is non-singular. That
is, $\bar{\varSigma}_{w_{n}}$ becomes nearly singular as $w_{n}$
grows.

\subsection{On the Rank Condition}

We start with the simple regression case where there is only one regressor
along with the intercept, $X=(1,X_{1})$. To simplify notation, let
\begin{equation}
\alpha(X)=\exp\left(X_{1}\right),\label{eq:alpha}
\end{equation}
where the intercept term is 0, and $X_{1}$ has a compact support
on $\left[\underline{u}_{x},\bar{u}_{x}\right]$ with $\bar{u}_{x}>\underline{u}_{x}$.
Note that \eqref{eq:alpha} is equivalent to 
\[
\alpha(X)=\exp\left(\theta_{0}^{*}+\theta_{1}^{*}\tilde{X}_{1}\right),
\]
where $\theta_{0}^{*}=\underline{u}_{x},\theta_{1}^{*}=\bar{u}_{x}-\underline{u}_{x}$
and $\tilde{X}_{1}=(X_{1}-\underline{u}_{x})\left/\left(\bar{u}_{x}-\underline{u}_{x}\right)\right.$,
and the support of $\tilde{X}_{1}$ is $[0,1]$. Thus, \eqref{eq:alpha}
is a normalization of a general linear index.

For illustration, we write 
\[
Z=\exp\left(X_{1}\right).
\]
We will show the properties of $Z$ first, then extend the results
to $X.$

\begin{assumption}\label{A:1} The support of $X_{1}$ is bounded,
that is, $-\infty<\underline{u}_{x}<\bar{u}_{x}<\infty$. The density
function of $X_{1}$ is bounded and bounded away from zero: $0<\underline{c}_{x}\leq f(x_{1})\leq\bar{c}_{x}<\infty.$

\end{assumption}

Denote $\underline{u}=\exp\left(\underline{u}_{x}\right)$ and $\bar{u}=\exp\left(\bar{u}_{x}\right)$,
then the support of $Z$ is $\left[\underline{u},\bar{u}\right]$,
and the density of $Z$ is bounded and bounded away from zero; specifically,
there exist constants such that 
\[
0<\underline{c}\leq f(z)\leq\bar{c}<\infty\textrm{ for }z\in[\underline{u},\bar{u}].
\]
For simplicity of analysis, we impose $L(y;x)=1$ in the formal concentration
results below in order to isolate the design-degeneracy mechanism.
When $L(y;x)$ is nonconstant, the same concentration mechanism must
be considered jointly with the second-order approximation bias; Remark
2 discusses this interaction using the expansion in \citet{WangTsai2009}.

\begin{theorem}\label{TH:1} Suppose Assumption \ref{A:1} holds
and $Y$ follows the conditional distribution in \eqref{eq:surviveCDF}
with $L=1$. Then, 
\[
f(z\mid Y>w)\leq\frac{\bar{c}}{\underline{c}}\frac{w^{-(z-\underline{u})}\log w}{1-w^{-(\bar{u}-\underline{u})}},\quad z\in[\underline{u},\bar{u}],
\]
and as $w\to\infty$, 
\begin{align*}
\mathbb{E}(Z\mid Y>w) & \to\underline{u}\textrm{ \textrm{ and}},\\
\operatorname{Var}(Z\mid Y>w) & \to0.
\end{align*}
\end{theorem}

Theorem \ref{TH:1} shows that $Z$ degenerates to its lower bound
as $w\to\infty$.

Assumption \ref{A:1} requires the density of $Z$ to be bounded away
from zero on the whole support, which is stronger than necessary.
From the proof of Theorem \ref{TH:1}, it is enough that $f$ is bounded
above and satisfies, for some constants $k\geq0$, $\delta>0$ and
$\underline{c}>0$, 
\[
f(z)\geq\underline{c}\left(z-\underline{u}\right)^{k},\qquad z\in\left[\underline{u},\underline{u}+\delta\right],
\]
which reduces to Assumption \ref{A:1} when $k=0$. Under this weaker
requirement the conclusions of Theorem \ref{TH:1} continue to hold,
with the density bound replaced by $f(z\mid Y>w)\leq Cw^{-(z-\underline{u})}\left(\log w\right)^{k+1}$
for some $C>0$; see Step 1 of the proof of Corollary \ref{Coro:2}
in the Appendix. This weaker condition matters for the multiple-regressor
case below, where the density of the relevant covariate goes to zero
at the lower endpoint of its support.

Note that $X_{1}=\log Z.$ The conditional density of $X_{1}$ takes
a similar form by the change of variable method. Lemma \ref{LE:1}
shows that the result applies to $X_{1}$ as well: 
\begin{align}
\mathbb{E}(X_{1} & \mid Y>w)\rightarrow\log\left(\underline{u}\right)=\underline{u}_{x}\textrm{ and}\nonumber \\
\textrm{Var}(X_{1} & \mid Y>w)\rightarrow0.\label{eq:var->0}
\end{align}

In the special case where $Z$ is uniformly distributed or equivalently
$\alpha(X)=\exp\left(X_{1}\right)$ is uniformly distributed, we are
able to derive the rate at which $\operatorname{Var}(Z\mid Y>w)$
converges to zero.

\begin{corollary}\label{Coro:1} Suppose the assumptions in Theorem
\ref{TH:1} hold and $Z$ is uniformly distributed on $[\underline{u},\bar{u}]$.
Then, as $w\to\infty$, 
\begin{align*}
\mathbb{E}(Z\mid Y>w) & =\underline{u}+\frac{1}{\log w}+o\!\left(\frac{1}{\log w}\right)\textrm{ \textrm{ and}},\\[6pt]
\operatorname{Var}(Z\mid Y>w) & =\frac{1}{(\log w)^{2}}+o\!\left(\frac{1}{(\log w)^{2}}\right).
\end{align*}
\end{corollary}

For $X_{1}=\log Z$, Lemma \ref{LE:2} shows 
\begin{equation}
\frac{1}{\bar{u}^{2}}\textrm{Var}\left(Z\mid Y>w\right)\leq\textrm{Var}\left(X_{1}\mid Y>w\right)\leq\frac{1}{\underline{u}^{2}}\textrm{Var}\left(Z\mid Y>w\right).\label{eq:varX1bound}
\end{equation}
Therefore, the minimum eigenvalue of $\bar{\varSigma}_{w_{n}}$, defined
in (\ref{eq:Sigma_w_bar}), is proportional to $\textrm{Var}\left(X_{1}\mid Y>w\right)$
and is at the rate of $1/(\log w)^{2}.$

Now consider the case with multiple regressors, $X=(1,X_{1},X_{2},\ldots,X_{p})$.
Suppose we are in the best case for the rank condition of the Gram
matrix: the regressors are mutually independent, $X_{1}\perp X_{2}\perp\cdots\perp X_{p}$,
and each $X_{j}$ satisfies the support and density conditions in
Assumption \ref{A:1}. In addition, assume 
\[
\alpha(X)=\exp\left(X_{1}+X_{2}+\cdots+X_{p}\right).
\]

Define $\tilde{X}_{1}=X_{1}+X_{2}+\cdots+X_{p}$ and $\underline{\tilde{u}}_{x_{1}}=\inf\tilde{X}_{1}$.
For $p\geq2$, $\tilde{X}_{1}$ does not satisfy Assumption \ref{A:1}:
the density of a sum of independent covariates whose densities are
bounded away from zero vanishes at the endpoints of its support, at
the rate $\left(s-\underline{\tilde{u}}_{x_{1}}\right)^{p-1}$ near
the lower endpoint. It does, however, satisfy the weaker requirement
stated after Theorem \ref{TH:1} with $k=p-1$. Applying that extension
of Theorem \ref{TH:1} together with the result in (\ref{eq:var->0})
yields:

\begin{corollary}\label{Coro:2} Suppose $X_{1},X_{2},\ldots,X_{p}$
satisfy Assumption \ref{A:1} and are mutually independent. Let $\underline{\tilde{u}}_{x_{1}}=\inf\tilde{X}_{1}$.
Then, as $w\to\infty$, 
\begin{align*}
\mathbb{E}(\tilde{X}_{1}\mid Y>w) & \to\underline{\tilde{u}}_{x_{1}}\textrm{\textrm{ and}},\\[6pt]
\operatorname{Var}(\tilde{X}_{1}\mid Y>w) & \to0.
\end{align*}
\end{corollary}

In other words, $(1,X_{1},X_{2},\ldots,X_{p})$ becomes nearly collinear
as $w\to\infty$, since $\tilde{X}_{1}=X_{1}+X_{2}+\cdots+X_{p}$
behaves like a constant. Thus, under fairly general conditions, the
population matrix $\bar{\varSigma}_{w}$ degenerates to a singular
matrix as $w\to\infty$.

The bounded-above support assumption is mainly to simplify the presentation.
We can generalize the result in Theorem \ref{TH:1} to the case where
$X_{1}$ is unbounded above, while retaining a finite lower endpoint.
To save notation, we impose conditions on $Z$ directly. The corresponding
conditions on $X_{1}$ follow from the relation $Z=\exp\left(X_{1}\right).$

\begin{corollary}\label{Coro:2-1} Suppose that $Y$ follows the
conditional distribution in (\ref{eq:surviveCDF}) with $L=1$. $Z$
takes values in $[\underline{u},\infty)$ with $\underline{u}>0$,
and $f(z)>\underline{c}>0$ for $z\in[\underline{u},\underline{u}+\delta]$
for some $\delta>0$. Also, $f(z)<\overline{c}<\infty$ for all $z$.
Then, 
\[
f(z\mid Y>w)\le\frac{\bar{c}}{\underline{c}}\frac{w^{-(z-\underline{u})}\log w}{1-w^{-\delta}}.
\]
In addition, suppose $\sup_{w>0}\mathbb{E}\left(Z^{2+\eta}\mid Y>w\right)<\infty$
for some $\eta>0$. Then $\mathbb{E}(Z\mid Y>w)\to\underline{u}\textrm{ \textrm{ and} }\textrm{Var}(Z\mid Y>w)\to0.$

\end{corollary}

Note that the additional conditions are rather mild for $Z$ with
support unbounded above. We only need $f\left(z\right)$ to be uniformly
bounded away from zero in a small neighborhood of $\underline{u},$
$f\left(z\right)$ uniformly bounded, and finite $\left(2+\eta\right)$-th
moment of $Z$ conditional on the event of $\left\{ Y>w\right\} .$
This finite moment condition is only needed to show the conditional
variance of $Z$ converges to zero and not required for the degenerate
density result. If one wishes to investigate the conditional variance,
it is standard to assume a slightly higher moment condition.

\subsection{The Asymptotics}

In this section, we provide additional conditions and a new proof
showing that the main result in \citet{WangTsai2009} is correct,
though with an extra condition.

We continue to assume $X=(1,X_{1},X_{2},\ldots,X_{p})$. Let 
\[
\bar{\varSigma}_{w_{n}}=\mathbb{E}\!\left(XX'|Y>w_{n}\right)
\]
denote the population counterpart of $\hat{\varSigma}_{w_{n}}$. Because
$X$ contains an intercept, the upper-left element of $\bar{\varSigma}_{w_{n}}$
equals one, and hence $\rho_{\max}(\bar{\varSigma}_{w_{n}})\geq1$.
Thus, the eigenvalues of $\bar{\varSigma}_{w_{n}}$ cannot all converge
to zero at a common rate. Some directions can keep order-one information
while other directions become weakly identified.

Define 
\[
\underline{\lambda}_{n}=\rho_{\min}\!\left(\bar{\varSigma}_{w_{n}}\right).
\]
We impose no common-rate restriction on the remaining eigenvalues.
Instead, the rate condition is stated directly in terms of the weakest
direction: 
\begin{equation}
n\Pr(Y>w_{n})\underline{\lambda}_{n}^{2}=n\Pr(Y>w_{n})\left[\rho_{\min}\!\left(\bar{\varSigma}_{w_{n}}\right)\right]^{2}\longrightarrow\infty.\label{eq:varying_rank}
\end{equation}
Since $n_{0}\left/\left(n\Pr(Y>w_{n})\right)\right.\overset{P}{\longrightarrow}1$,
condition (\ref{eq:varying_rank}) is equivalently written as 
\[
n_{0}\left[\rho_{\min}\!\left(\bar{\varSigma}_{w_{n}}\right)\right]^{2}\overset{P}{\longrightarrow}\infty.
\]
Under the moment condition below, $\rho_{\max}(\bar{\varSigma}_{w_{n}})$
is uniformly bounded and is bounded below by one, because of the intercept.

To simplify the analysis, we assume $L(y;x)=1$, so that we do not
need to account for the bias term in \citet{WangTsai2009}. The results
are qualitatively unchanged if $L(y;x)$ is not constant.

\begin{theorem}\label{TH:2} Suppose $L=1$. $\mathbb{E}\!\left[\|X\|^{2+\delta}\mid Y>w_{n}\right]$
is uniformly bounded for some $\delta\geq2$. Condition (\ref{eq:varying_rank})
holds. Let $(x_{i},y_{i}),i=1,2,\ldots,n$, be i.i.d. across $i$.
Then, using the sample matrix $\hat{\varSigma}_{w_{n}}$, we have
\[
\sqrt{n_{0}}\,\hat{\varSigma}_{w_{n}}^{1/2}(\hat{\theta}-\theta^{*})\;\overset{d}{\longrightarrow}\;N(0,\mathbb{I}_{p+1}).
\]
\end{theorem}

Theorem \ref{TH:2} has the same form as the main theorem in \citet{WangTsai2009},
but with two key differences. First, the normalization is matrix-valued
and therefore allows different parameter directions to converge at
different rates. To see this, let $\lambda_{jn}$ and $v_{jn}$ denote
an eigenvalue and the corresponding unit eigenvector of $\bar{\varSigma}_{w_{n}}$.
Theorem \ref{TH:2} implies that the rate in direction $v_{jn}$ is
$\sqrt{n_{0}\lambda_{jn}}$. Thus, a direction with $\lambda_{jn}$
bounded away from zero retains the usual $\sqrt{n_{0}}$ rate, whereas
the weakest direction has rate $\sqrt{n_{0}\underline{\lambda}_{n}}$.
In particular, 
\[
\|\hat{\theta}-\theta^{*}\|=O_{P}\!\left(\frac{1}{\sqrt{n_{0}\underline{\lambda}_{n}}}\right)
\]
is the slowest-direction bound, but the estimator need not have a
common scalar rate. Second, condition (\ref{eq:varying_rank}) requires
$\sqrt{n\Pr(Y>w_{n})}\,\underline{\lambda}_{n}\to\infty$. This condition
allows the smallest eigenvalue to converge to zero, provided that
it does not do so too quickly relative to the effective tail sample
size. It also allows the other eigenvalues to stay of order one or
to converge to zero at different rates. In practice, there is no need
to estimate a scalar degeneracy rate; inference can be conducted using
the matrix-normalized asymptotic distribution in Theorem \ref{TH:2}
directly.

\begin{remark}[On condition (\ref{eq:varying_rank})] \emph{Condition
(\ref{eq:varying_rank}) can be written as 
\[
\sqrt{n\Pr(Y>w_{n})}\,\underline{\lambda}_{n}\to\infty.
\]
It ensures that the sampling error stays negligible after we standardize
by the inverse population Gram matrix. In particular, the proof below
establishes 
\[
\left\Vert \bar{\varSigma}_{w_{n}}^{-1/2}(\hat{\varSigma}_{w_{n}}-\bar{\varSigma}_{w_{n}})\bar{\varSigma}_{w_{n}}^{-1/2}\right\Vert =O_{P}\!\left(\frac{1}{\sqrt{n_{0}}\,\underline{\lambda}_{n}}\right)=o_{P}(1).
\]
The same condition is sufficient for the Lyapunov argument used in
the score central limit theorem. Since $\frac{n_{0}}{n\Pr(Y>w_{n})}\overset{P}{\longrightarrow}1$,
it is equivalently read as $\sqrt{n_{0}}\,\underline{\lambda}_{n}\overset{P}{\longrightarrow}\infty$.
In the uniform example of Corollary \ref{Coro:1}, $\underline{\lambda}_{n}\propto(\log w_{n})^{-2}$.
Moreover, standard calculations yield $\Pr(Y>w_{n})\propto1/[(w_{n})^{\underline{u}}\log w_{n}]$.
Therefore, condition (\ref{eq:varying_rank}) becomes 
\[
\frac{(\log w_{n})^{5}(w_{n})^{\underline{u}}}{n}\to0.
\]
Similar conditions can be derived for other specifications.}

\end{remark}

\begin{remark}[Bias and Variance Trade-off under Degeneracy]\emph{Suppose
$L$ follows the second-order expansion in Wang and Tsai (2009): 
\[
\bar{F}(y\mid x)=y^{-\alpha(x)}L(y;x),\qquad L(y;x)=c_{0}(x)+c_{1}(x)y^{-\beta(x)}+o\!\left(y^{-\beta(x)}\right),
\]
where $c_{0}(x)>0$ and $\beta(x)>0$. Under an asymptotically degenerate
population Gram matrix $\bar{\Sigma}_{w_{n}}$, both the bias and
variance are affected by the ill-conditioning of $\bar{\Sigma}_{w_{n}}$.
In particular, from $\mathbf{Q}_{n1}-\mathbf{Q}_{n2}$ in Appendix
B and the expansion of $\sqrt{n_{0}}\hat{\boldsymbol{\gamma}}_{n}$
in Appendix C of Wang and Tsai (2009), the bias term is inflated at
the order $\bar{\Sigma}_{w_{n}}^{-1}$, while the variance is inflated
at the same matrix order. Consequently, the standard deviation is
inflated at the order $\bar{\Sigma}_{w_{n}}^{-1/2}$. Thus, in a weak
eigendirection for which the second-order bias has a nonzero component,
inversion by the Gram matrix can amplify that bias at order $1/\lambda$,
whereas the corresponding standard deviation is amplified at order
$1/\sqrt{\lambda}$. This directional comparison does not apply when
the bias component in that eigendirection is zero, and by itself it
does not determine how the optimal threshold should move. Increasing
$w_{n}$ reduces the second-order approximation bias, but at the same
time reduces the effective tail sample size $n_{0}$ and usually makes
the design degeneracy worse through a smaller minimum eigenvalue $\underline{\lambda}_{n}$.
Therefore, the optimal threshold depends jointly on the rate of the
approximation bias, the effective tail sample size, and the eigenvalues
of the Gram matrix. This point can be seen more formally from condition
(C.3) in Wang and Tsai (2009), which requires the bias and the standard
deviation to be of the same order. Specifically, 
\begin{equation}
\frac{n}{\sqrt{n_{0}}}\mathbb{E}\left[\bar{\Sigma}_{w_{n}}^{-1/2}X\frac{c_{1}(X)\beta(X)}{\alpha(X)+\beta(X)}w_{n}^{-\alpha(X)-\beta(X)}\right]\rightarrow\mathbf{h},\label{eq:balance_bias_std}
\end{equation}
for some nonzero constant vector $\mathbf{h}$. Thus, degeneracy enters
the bias--variance balance through the full matrix $\bar{\Sigma}_{w_{n}}^{-1/2}$,
but condition (\ref{eq:balance_bias_std}) by itself does not imply
a monotone conclusion for the optimal choice of $w_{n}$.}

\end{remark}

\begin{remark}[A Practical Way to Choose $w_{n}$] \emph{Wang and
Tsai (2009, Section 2.3) propose a discrepancy measure for threshold
selection based on a distributional approximation. Under a correct
threshold $w_{n}$, the transformed variable 
\[
\hat{U}_{i}=\exp\{-\exp(x_{i}'\hat{\theta})\log(y_{i}/w_{n})\},\quad y_{i}>w_{n},
\]
is approximately uniformly distributed on $[0,1]$. When the sample
fraction is too large, this approximation becomes inaccurate; when
it is too small, the estimator suffers from high variance. Based on
this observation, Wang and Tsai (2009) select $w_{n}$ by minimizing
the discrepancy between the empirical distribution of $\{\hat{U}_{i}:y_{i}>w_{n}\}$
and the uniform distribution: 
\[
\hat{D}(w_{n})=n^{-1}\sum_{i=1}^{n}\left\{ \hat{U}_{i}-\hat{F}_{n}(\hat{U}_{i})\right\} ^{2}I(y_{i}>w_{n}),
\]
where $\hat{F}_{n}(\cdot)$ is the empirical distribution of $\hat{U}_{i}$,
and $w_{n}$ is chosen to minimize $\hat{D}(w_{n})$. This discrepancy
measure provides a practical data-driven procedure for threshold selection.
However, its data-driven nature alone does not establish that it automatically
adapts to the degeneracy of the Gram matrix. Showing such an adaptation
would require a separate analysis, which we do not carry out here.
We therefore view the discrepancy measure as a practical threshold-selection
procedure rather than a formal solution to the degeneracy problem.}

\end{remark}

\subsection{Semi/Non-parametric Tail Index Regressions}

The problem is more serious in the semiparametric case, especially
for the nonparametric component. We first derive the conditional density
of $X$ in the tail, and then present the results regarding the semiparametric
regression.

\subsubsection*{Density of $X$ on the Tail of $Y$}

For convenience, suppose that 
\begin{equation}
\alpha(X)=X_{1},\label{eq:alphax_semiparametric}
\end{equation}
and we wish to estimate $\alpha(X)$ nonparametrically. The support
of $X_{1}$ is $[\underline{u}_{x},\bar{u}_{x}]$ with $\underline{u}_{x}>0$.

Assume the density of $X_{1}$ is bounded and bounded away from zero.
This is the most favorable scenario for nonparametric estimation.
Theorem \ref{TH:1} implies that 
\[
f(x_{1}\mid Y>w)\;\leq\;C\frac{w^{-(x_{1}-\underline{u}_{x})}\log w}{1-w^{-(\bar{u}_{x}-\underline{u}_{x})}},\quad x_{1}\in[\underline{u}_{x},\bar{u}_{x}],
\]
for some $C>0.$ This shows that the conditional density of $X_{1}$
converges to zero for all $x_{1}\in(\underline{u}_{x},\bar{u}_{x}]$,
i.e., for all values in the support of $X_{1}$ except the minimum,
even under the best-case scenario. The speed of decay becomes faster
as $x_{1}$ moves farther away from its minimum. The problem can become
worse when there is bias, that is, when $L\left(y;x\right)$ is not
constant. The bias term will dominate more easily as $x_{1}$ moves
away from its minimum.

The discussion under condition \eqref{eq:alphax_semiparametric} can
be generalized. Suppose we have a general nonlinear $\alpha(X)$,
and $X$ is a scalar. Define 
\[
Z=\alpha(X).
\]
Assume $\alpha(x)$ is continuously differentiable, with $\alpha'(x)$
bounded and bounded away from zero, and the density of $X$ bounded
and bounded away from zero. For any value of $z$, there exist only
finitely many $x$ such that $\alpha(x)=z$. Then $Z$ also has bounded
support, and 
\begin{equation}
f(z)=\sum_{\alpha(x)=z}\frac{f(x)}{|\alpha'(x)|},\label{eq:fzfx}
\end{equation}
which is well defined for all $z$ in the support of $Z$. It is easy
to see that $f(z)$ is also bounded and bounded away from zero. If
we denote the support of $Z$ as $[\underline{u},\bar{u}]$, then
applying the previous result we obtain 
\begin{equation}
f(z\mid Y>w)\;\leq\;C\,\frac{w^{-(z-\underline{u})}\log w}{1-w^{-(\bar{u}-\underline{u})}},\quad z\in[\underline{u},\bar{u}],\label{eq:fz|y}
\end{equation}
for some positive constant $C$. The inequality above, together with
\eqref{eq:fzfx}, implies that $f(x\mid Y>w)$ converges to zero whenever
$\alpha(x)\neq\underline{u}$, with faster decay the further $\alpha(x)$
lies above its minimum.

\subsubsection*{Semiparametric Regression}

Suppose $X=\left[1,X_{1},X_{2}\right]$. The semiparametric framework
in \citet{LiLengYou2022} can be written as 
\[
\alpha(X)=\exp\left[\theta_{0}^{*}+\theta_{1}^{*}X_{1}+g^{*}\left(X_{2}\right)\right],
\]
where $g^{*}$ is some unknown function. To estimate the nonparametric
component together with the parametric component, the first step in
\citet{LiLengYou2022} approximates the nonparametric part by sieves,
so that both components can be estimated at the same time. That is,
\[
\min_{\theta,g_{n}}\sum_{i=1}^{n}\Bigl\{\exp\left[\theta_{0}+\theta_{1}x_{1i}+g_{n}\left(x_{2i}\right)\right]\,\log\!\left(\frac{y_{i}}{w_{n}}\right)-\theta_{0}-\theta_{1}x_{1i}-g_{n}\left(x_{2i}\right)\Bigr\}\,I(y_{i}>w_{n}),
\]
where $g_{n}$ is an approximation of $g^{*}$ using sieves (e.g.,
B-splines). Thus, the objective above has the same form as a parametric
tail index regression.

This may create problems for the nonparametric component, $g^{*}\left(X_{2}\right)$,
due to the degenerate density function in (\ref{eq:fz|y}). Specifically,
most of the extreme observations are likely to be concentrated around
the point where the minimum of $\alpha$ is attained. Consequently,
$g^{*}$ is weakly identified. However, it is difficult to compute
the convergence rate at each point, because sieve estimators are global
and the conditional density degenerates at different rates across
$x_{2}$.

\subsubsection*{Nonparametric Regression}

The nonparametric tail index regression in \citet{deHaan03072021}
is less affected by the issue we identify, because it relies only
on local data for estimation and $\alpha(x)$ behaves approximately
like a constant locally. In addition, they use order statistics to
decide which observations to include in the local estimation. Clearly,
order statistics reflect the value of the index.

To further clarify the role of degeneracy in nonparametric estimation
based on samples $\{Y>w\}$, we provide a formal analysis in Appendix
\ref{APP:B}. In particular, we consider a simple estimator based
on degree-0 B-splines (equivalently, a uniform kernel estimator) using
exceedance samples $\{Y>w\}$ for all $x$, rather than order statistics.
We derive a pointwise convergence-rate bound and show explicitly how
the conditional density $f(x\mid Y>w)$ determines the effective sample
size in each local region. The results show that, as $f(x\mid Y>w)$
becomes small in a neighborhood of $x$, the convergence rate deteriorates
accordingly.

\section{A Comparison with Extremal Quantile Regression}

\label{sec:compare}

The case of right-tail behavior of $Y$ conditional on $X$ with unbounded
$Y$ corresponds to the type 2 tail in \citet{Chernozhukov2005}.
We focus on the type-2, unbounded-tail case. Extensions to bounded
tails require separate treatment of covariate-dependent endpoints
and are not pursued here.

\subsection{The Model}

It is assumed in \citet{Chernozhukov2005} that 
\begin{equation}
Y=\beta\left(X\right)+U,\label{eq:quantile_model}
\end{equation}
and 
\begin{equation}
\bar{F}_{U}(u\mid x)=1-F_{U}\left(u|x\right)=k\left(x\right)\cdot u^{-\alpha}L(u),\label{eq:surviveCDF1}
\end{equation}
where $\alpha$ is a positive constant, $\beta\left(\cdot\right)$
and $k\left(\cdot\right)$ are the location and scale functions, respectively;
$F_{U}\left(u|x\right)$ is the cumulative distribution function of
$U$ conditional on $X=x$; and $L(y)$ is a slowly varying function
that satisfies $L(yt)/L(y)\to1$ for any $t>0$ as $y\to\infty$.

From (\ref{eq:surviveCDF}) and (\ref{eq:surviveCDF1}), the key difference
is where $X$ enters. $k\left(X\right)$ in (\ref{eq:surviveCDF1})
affects the cumulative distribution function (CDF) as a scale function,
and the tail index $\alpha$ does not depend on $X$. On the other
hand, $X$ enters the tail index as $\alpha(X)$ in (\ref{eq:surviveCDF}).
Clearly, $X$ has a stronger effect on the tail distribution in (\ref{eq:surviveCDF})
than in (\ref{eq:surviveCDF1}). For example, changing $k\left(x\right)=2$
to $k\left(x\right)=4$ doubles the density at the tail while the
tail decays to zero at the same rate, but changing $\alpha(x)$ from
2 to 4 leads to $y^{-2}$ becoming $y^{-4}$, which are completely
different tail behaviors. This observation provides intuition for
the different behaviors of $X$ conditional on $Y$ in the tails under
these two frameworks.

\subsection{The Quantiles}

We present the implications of the tail CDF for the extremal $\tau$-th
quantile of $Y$ ($\tau$ close to 1). Since $L$ is slowly varying
at the right tail, we set $L\left(y;x\right)$ and $L\left(y\right)$
equal to 1 as an approximation to simplify the analysis.

For \citet{Chernozhukov2005}, (\ref{eq:surviveCDF1}) implies that
\begin{equation}
Q_{Y}\left(\tau|X\right)\approx\beta\left(X\right)+\left[k\left(X\right)\right]^{1/\alpha}\left(1-\tau\right)^{-1/\alpha},\label{eq:quantile1}
\end{equation}
where $Q_{Y}\left(\tau|x\right)$ denotes the $\tau$-th quantile
of $Y$ conditional on $X=x$, and ``$\approx$'' holds by setting
$L\left(y\right)=1$.

Similarly, (\ref{eq:surviveCDF}) in \citet{WangTsai2009} implies
\begin{equation}
Q_{Y}\left(\tau|X\right)\approx\left(1-\tau\right)^{-1/\alpha(X)}.\label{eq:quantile2}
\end{equation}

The expressions in (\ref{eq:quantile1}) and (\ref{eq:quantile2})
confirm the observation in the previous section: $X$ has a stronger
effect on the tail behavior of $Y$ in the framework of \citet{WangTsai2009}
than in \citet{Chernozhukov2005}.

\subsection{Different Behaviors of $X$ on the Tail of $Y$}

\label{sec:3}

Assuming $L(u)=1$, (\ref{eq:quantile_model}) and (\ref{eq:surviveCDF1})
are equivalent to 
\begin{equation}
Y=\beta\left(X\right)+k\left(X\right)^{1/\alpha}\tilde{U},\textrm{ and }\bar{F}_{\tilde{U}}(\tilde{u}\mid x)=\tilde{u}^{-\alpha}.\label{eq:extremal_rewrite}
\end{equation}
We simply write $\beta\left(X\right)$ as $X_{1}$ and $k\left(X\right)^{1/\alpha}$
as $X_{2}$, both of which are scalars. Given the above, the following
representation is quite general: 
\begin{equation}
Y=X_{1}+X_{2}\cdot\tilde{U}\label{eq:quantile_equivalent}
\end{equation}
with 
\begin{equation}
X_{2}>0\textrm{ and }\bar{F}_{\tilde{U}}\left(\tilde{u}|X_{1},X_{2}\right)=\tilde{u}^{-\alpha},\alpha>1.\label{eq:equivalent_condition}
\end{equation}
We have the following result for the conditional density $f\left(X_{1},X_{2}|Y>w\right).$

\begin{theorem}\label{TH:density_quantile} Suppose the model in
(\ref{eq:quantile_equivalent}) with condition (\ref{eq:equivalent_condition})
holds. In addition, $\left(X_{1},X_{2}\right)$ takes values in $\left[\underline{u}_{x_{1}},\bar{u}_{x_{1}}\right]\times\left[\underline{u}_{x_{2}},\bar{u}_{x_{2}}\right]$,
$\underline{u}_{x_{2}}>0$, and the density of $\left(X_{1},X_{2}\right)$
is uniformly bounded and bounded away from zero. Then, uniformly over
$\left(x_{1},x_{2}\right)\in\left[\underline{u}_{x_{1}},\bar{u}_{x_{1}}\right]\times\left[\underline{u}_{x_{2}},\bar{u}_{x_{2}}\right]$,
\begin{equation}
f\left(x_{1},x_{2}|Y>w\right)\rightarrow\frac{x_{2}^{\alpha}f\left(x_{1},x_{2}\right)}{\int_{\underline{u}_{x_{2}}}^{\bar{u}_{x_{2}}}\int_{\underline{u}_{x_{1}}}^{\bar{u}_{x_{1}}}s_{2}^{\alpha}f\left(s_{1},s_{2}\right)\textrm{d}s_{1}\textrm{d}s_{2}}\textrm{ as }w\rightarrow\infty.\label{eq:conditional_den_x1x2|w}
\end{equation}
In particular, if $\left(X_{1},X_{2}\right)$ is uniformly distributed
on $\left[\underline{u}_{x_{1}},\bar{u}_{x_{1}}\right]\times\left[\underline{u}_{x_{2}},\bar{u}_{x_{2}}\right]$,
the limit in (\ref{eq:conditional_den_x1x2|w}) equals 
\begin{equation}
\frac{\left(\alpha+1\right)x_{2}^{\alpha}}{\left(\bar{u}_{x_{1}}-\underline{u}_{x_{1}}\right)\left(\bar{u}_{x_{2}}^{\alpha+1}-\underline{u}_{x_{2}}^{\alpha+1}\right)}.\label{eq:conditional_den_uniform}
\end{equation}
\end{theorem}

The limit in (\ref{eq:conditional_den_x1x2|w}) is the density of
$\left(X_{1},X_{2}\right)$ tilted by the factor $x_{2}^{\alpha}$.
The tilting shifts mass toward large values of $X_{2}$, or equivalently
toward large values of $k\left(X\right)$, which fits the intuition;
in the uniform case (\ref{eq:conditional_den_uniform}), the limiting
density is maximized at $x_{2}=\bar{u}_{x_{2}}$. Clearly, $\left(X_{1},X_{2}\right)|Y>w$
does not degenerate as $w\rightarrow\infty$, because the limit in
(\ref{eq:conditional_den_x1x2|w}) is a fixed density that is bounded
away from zero on the support of $\left(X_{1},X_{2}\right)$. This
is in clear contrast to the implication in (\ref{eq:fz|y}) that the
density converges to zero except at the point where $\alpha$ reaches
its minimum. We therefore conclude that the rank condition issue does
not exist for the extremal quantile regression framework. The intuition
can be seen from the previous two subsections: $X$ has a much stronger
impact on the tail behavior of $Y$ in the tail index regression framework
than in the extremal quantile regression framework.

\section{Numerical Verification}

\label{sec:Simulations}

\subsection{Data Generating Processes}

We check the rank condition for these two frameworks by simulations.
We consider three data-generating processes (DGPs). In the first DGP,
$\alpha\left(X\right)$ has a single global minimum in the tail index
regression setting, and $k\left(X\right)$ in (\ref{eq:surviveCDF1})
has a single global maximum in the extremal quantile regression setting.
We refer to this design as DGP1M, where 1M stands for ``one mode''.
In the second DGP, $\alpha\left(X\right)$ has four global minima,
and $k\left(X\right)$ has four global maxima. We refer to this design
as DGP4M. The second DGP is designed to describe situations where
extreme risk is high in more than one period. For the final DGP, we
extend DGP1M to include two covariates $X$, and refer to this specification
as DGP2X1M.

Specifically, the tail index regression setting in DGP1M follows 
\[
X\sim\textrm{Uniform}\left(\left[0,1\right]\right),
\]
and $Y$ follows the following CDF given $X=x$: 
\begin{align}
F\left(y|X=x\right) & =\left\{ \begin{array}{cc}
1-y^{-1.5-10x} & \textrm{ for }y\geq1,\\
0 & \textrm{ for }y<1.
\end{array}\right.\label{eq:FinDGP1M}
\end{align}
We label this design as ``DGP1M-Tail-Index''. In the extremal quantile
regression setting, $(Y,X)$ are generated according to 
\begin{align}
X & \sim\textrm{Uniform}\left(\left[0,1\right]\right),\nonumber \\
Y & =X+\left(11.5-10X\right)U,\textrm{ with }U\sim\left|\textrm{Student }t\left(4\right)\right|.\label{eq:YinDGP1M}
\end{align}
We correspondingly label this design as ``DGP1M-Extremal-Quantile''.
The mode of $f\left(X|Y>w\right)$ is $X=0$ for both settings.

In DGP4M, the tail index regression setting follows 
\[
X\sim\textrm{Uniform}\left(\left[0,1\right]\right),
\]
and $Y$ follows the following CDF given $X=x$: 
\begin{align*}
F\left(y|X=x\right) & =\left\{ \begin{array}{cc}
1-y^{-6.5+5\cos\left(20x\right)} & \textrm{ for }y\geq1,\\
0 & \textrm{ for }y<1.
\end{array}\right.
\end{align*}
The extremal quantile regression setting follows 
\begin{align*}
X & \sim\textrm{Uniform}\left(\left[0,1\right]\right),\\
Y & =X+\left[6.5+5\cos\left(20X\right)\right]U,\textrm{ with }U\sim\left|\textrm{Student }t\left(4\right)\right|.
\end{align*}
The modes of $f\left(X|Y>w\right)$ are $X=0,\pi/10,\pi/5,3\pi/10$
for both settings. We similarly label these two designs as ``DGP4M-Tail-Index''
and ``DGP4M-Extremal-Quantile''.

In DGP2X1M, we consider two independent covariates $X_{1}$ and $X_{2}$
such that 
\[
X_{1},X_{2}\sim\text{Uniform}([0,1]).
\]
The data-generating process follows DGP1M, except that we replace
$-1.5-10x$ in (\ref{eq:FinDGP1M}) with $-1.5-5(x_{1}+x_{2})$, and
replace $11.5-10X$ in (\ref{eq:YinDGP1M}) with $11.5-5(X_{1}+X_{2})$.
The mode of $f(X\mid Y>w)$ is attained at $X_{1}=X_{2}=0$ in both
settings. We refer to these two settings as ``DGP2X1M-Tail-Index''
and ``DGP2X1M-Extremal-Quantile,'' respectively.

\subsection{Tuning Parameter and Other Details}

Although $f\left(X|Y>w\right)$ can be computed analytically, the
derivation is long. We estimate the density of 
\[
f\left(X\mid Y>Q_{\tau}\left(Y\right)\right),\textrm{ for }\tau=0.9,0.95,0.99,0.995,
\]
for DGP1M and DGP4M. Here, $Q_{\tau}\left(Y\right)$ is estimated
by the $\tau$-th quantile of the sampled $Y$, and $f$ is obtained
using a histogram density estimator. Specifically, we define bins
\[
\mathbb{B}_{j}=\left(\left(j-1\right)h,jh\right],\quad j=1,2,\ldots,\left\lfloor 1/h\right\rfloor ,
\]
and 
\[
\hat{f}_{h}\left(x\mid Y>Q_{\tau}\left(Y\right)\right)=\frac{1}{n\left(1-\tau\right)h}\sum_{i:y_{i}>Q_{\tau}\left(Y\right)}1\left(x_{i}\in\mathbb{B}_{j}\right),\textrm{ for }x\in\mathbb{B}_{j},
\]
where $n$ is the number of observations. We set $h=0.01$.

For DGP1M, we report the variance of $X$ conditional on $Y>Q_{\tau}\left(Y\right)$.
For DGP2X1M, let $Z=(1,X_{1},X_{2})'$. To assess the full Gram matrix
directly, we report 
\[
\frac{\rho_{\min}\!\left(\widehat{\mathbb{E}}[ZZ'\mid Y>Q_{\tau}(Y)]\right)}{\rho_{\min}\!\left(\widehat{\mathbb{E}}[ZZ']\right)}.
\]
For DGP4M, instead of reporting $\textrm{Var}\left(X\mid Y>Q_{\tau}\left(Y\right)\right)$,
we report 
\begin{align}
 & \textrm{Var}_{4\textrm{M}}\left(X\mid Y>Q_{\tau}\left(Y\right)\right)\nonumber \\
\equiv\; & \textrm{Var}\left(X\mid Y>Q_{\tau}\left(Y\right),X<\frac{\pi}{20}\right)+\textrm{Var}\left(X\mid Y>Q_{\tau}\left(Y\right),\frac{\pi}{20}\leq X<\frac{3\pi}{20}\right)\nonumber \\
 & +\textrm{Var}\left(X\mid Y>Q_{\tau}\left(Y\right),\frac{3\pi}{20}\leq X<\frac{\pi}{4}\right)+\textrm{Var}\left(X\mid Y>Q_{\tau}\left(Y\right),X\geq\frac{\pi}{4}\right).\label{eq:var4M}
\end{align}
We use this modification because, under DGP4M, the distribution of
$X$ is concentrated around four modes at $X=0,\pi/10,\pi/5,3\pi/10$.
As a result, $\textrm{Var}\left(X\mid Y>Q_{\tau}\left(Y\right)\right)$
fails to capture the local variation around each mode. The cutoffs
are chosen as the midpoints between adjacent modes so that each term
in (\ref{eq:var4M}) measures the variation of $X$ in a neighborhood
of a single mode.

\subsection{Results}

\begin{figure}
\centering\caption{DGP1M}
\includegraphics[width=1\textwidth]{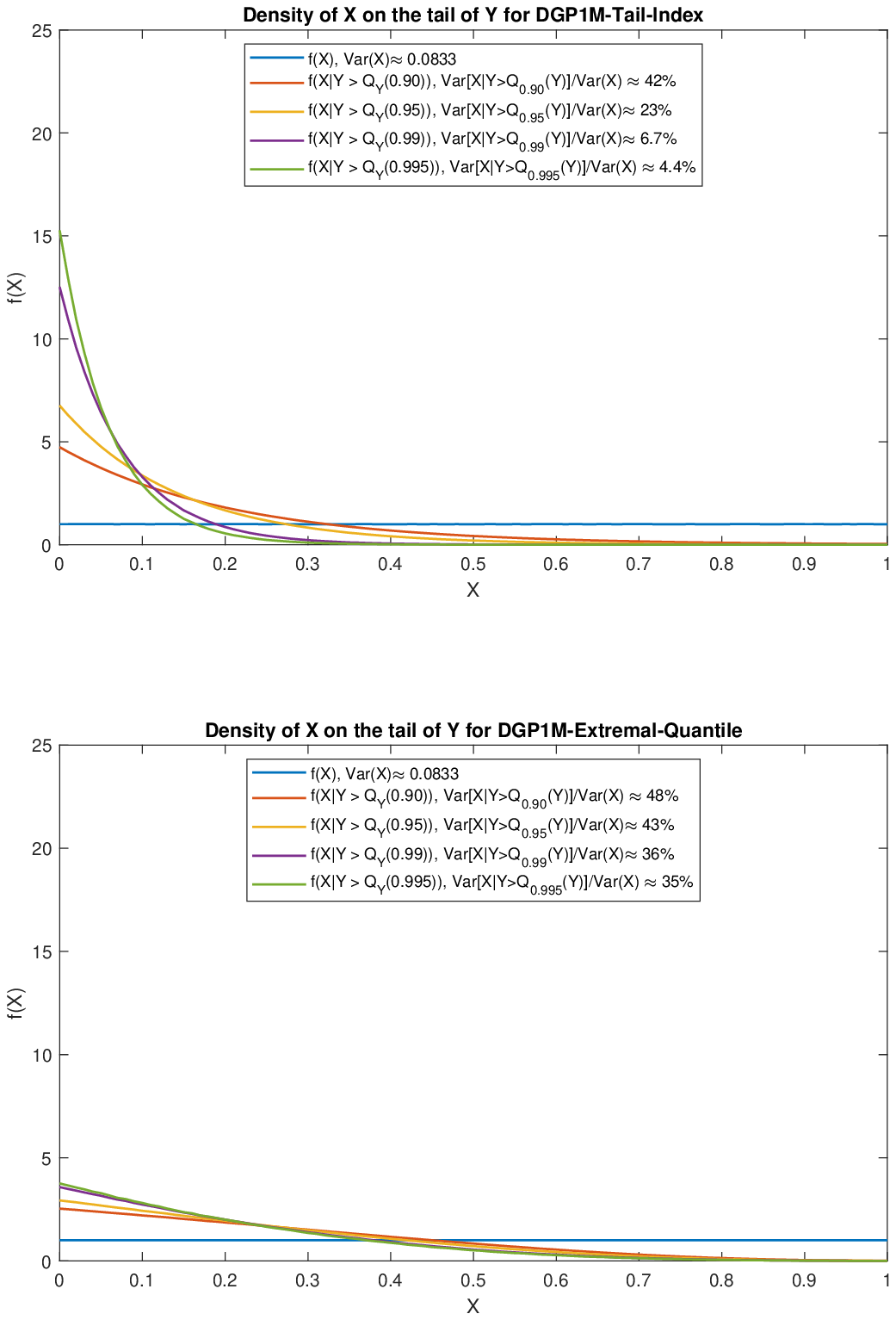} \label{fig:DPG1M} 
\end{figure}

\begin{figure}
\centering \caption{DGP4M}
\includegraphics[width=1\textwidth]{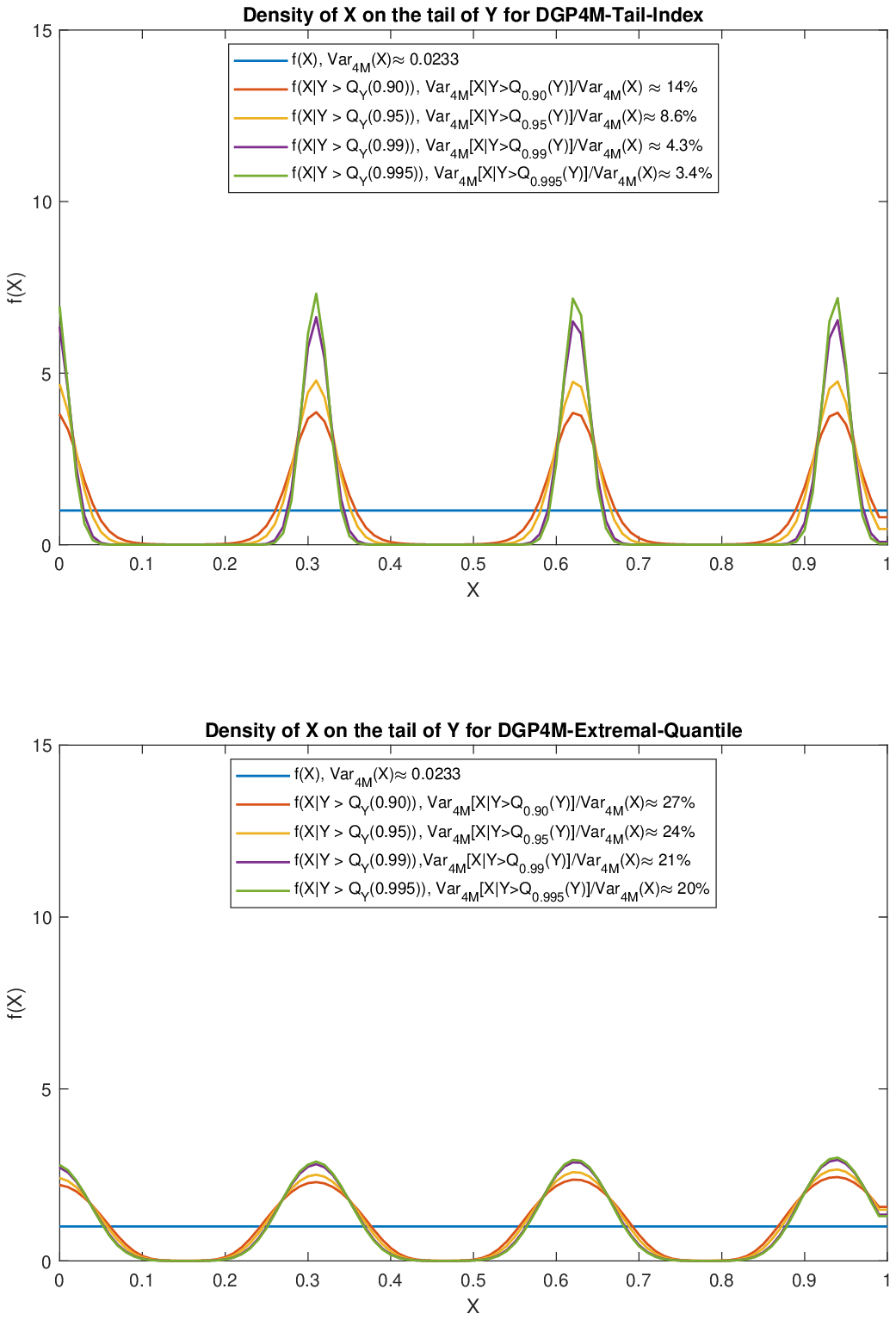} \label{fig:DPG4M} 
\end{figure}

To better visualize the results, we first draw $5\times10^{8}$ i.i.d.\ observations
for DGP1M and DGP4M and $10^{8}$ i.i.d.\ observations for DGP2X1M.
We report the estimated densities of $f\left(X\mid Y>Q_{\tau}\left(Y\right)\right)$
for DGP1M and DGP4M in Figures \ref{fig:DPG1M} and \ref{fig:DPG4M},
respectively. We report the variance ratios of the conditional variance
to the unconditional variance for DGP1M and DGP4M, and the minimum-eigenvalue
ratio of the Gram matrix for DGP2X1M. Since the number of sampled
observations is extremely large, these results are close to their
population counterparts. As a result, we conduct each simulation only
once.

To evaluate these measures in finite samples, we estimate them with
$n=1000$ and $n=5000$, based on 1000 replications. We report the
mean, standard deviation, and the 25th, 50th, and 75th percentiles
across the replications. The results are presented in Tables 1, 2,
and 3.

We first discuss the close-to-population result. As seen in Figures
\ref{fig:DPG1M} and \ref{fig:DPG4M}, the different behaviors of
$X$ conditional on the tail of $Y$ under the varying tail index
(tail index regression framework) and the constant tail index (extremal
quantile regression framework) are evident. The density of $X$ quickly
degenerates as $\tau$ increases toward one (or, in plain words, as
$Y$ becomes more extreme) for DGP1M-Tail-Index and DGP4M-Tail-Index.
This behavior is reflected in the conditional variance of $X$, which
decreases rapidly as $\tau$ increases: the conditional variance of
$X$ is only about 4\% of the unconditional variance of $X$ at $\tau=0.995$
for both DGP1M-Tail-Index and DGP4M-Tail-Index. In contrast, for DGP1M-Extremal-Quantile
and DGP4M-Extremal-Quantile, $X$ also concentrates more around the
modes, the conditional density of $X$ stabilizes and does not degenerate.
Moreover, the variance of $X$ conditional on $Y>Q_{0.995}\left(Y\right)$
is about 35\% and 20\% of the unconditional variance of $X$ for DGP1M-Extremal-Quantile
and DGP4M-Extremal-Quantile, respectively, changing little from the
variance conditional on $Y>Q_{0.99}\left(Y\right)$.

For DGP2X1M, we directly examine the minimum eigenvalue of the full
Gram matrix. For DGP2X1M-Tail-Index, the ratios 
\[
\frac{\rho_{\min}\!\left(\widehat{\mathbb{E}}[ZZ'\mid Y>Q_{\tau}(Y)]\right)}{\rho_{\min}\!\left(\widehat{\mathbb{E}}[ZZ']\right)}
\]
at $\tau=0.90,0.95,0.99,0.995,$ and $0.999$ are 1.010, 0.916, 0.590,
0.454, and 0.236, respectively. Thus, the minimum eigenvalue declines
substantially as the threshold increases, which shows directly that
the full Gram matrix deteriorates. For DGP2X1M-Extremal-Quantile,
the corresponding ratios are 0.849, 0.833, 0.814, 0.809, and 0.802,
respectively, indicating a clear pattern of stabilization.

\begin{table}[htbp]
\label{Tab:1} \centering \caption{DGP1M Variance ratios in small samples}
\begin{tabular}{c|ccccc}
\hline 
 & Mean  & Std  & 25\%  & 50\%  & 75\% \tabularnewline
\hline 
 & \multicolumn{5}{c}{Tail-Index $n=1000$}\tabularnewline
\hline 
$\text{Var}[X|Y>Q_{0.90}(Y)]/\text{Var}(X)$  & 0.420  & 0.094  & 0.354  & 0.416  & 0.477 \tabularnewline
$\text{Var}[X|Y>Q_{0.95}(Y)]/\text{Var}(X)$  & 0.241  & 0.091  & 0.177  & 0.226  & 0.296 \tabularnewline
$\text{Var}[X|Y>Q_{0.99}(Y)]/\text{Var}(X)$  & 0.074  & 0.075  & 0.029  & 0.052  & 0.092 \tabularnewline
\hline 
 & \multicolumn{5}{c}{Tail-Index $n=5000$}\tabularnewline
\hline 
$\text{Var}[X|Y>Q_{0.90}(Y)]/\text{Var}(X)$  & 0.421  & 0.041  & 0.393  & 0.421  & 0.449 \tabularnewline
$\text{Var}[X|Y>Q_{0.95}(Y)]/\text{Var}(X)$  & 0.235  & 0.041  & 0.208  & 0.231  & 0.259 \tabularnewline
$\text{Var}[X|Y>Q_{0.99}(Y)]/\text{Var}(X)$  & 0.066  & 0.027  & 0.047  & 0.061  & 0.080 \tabularnewline
\hline 
 & \multicolumn{5}{c}{Extremal-Quantile $n=1000$}\tabularnewline
\hline 
$\text{Var}[X|Y>Q_{0.90}(Y)]/\text{Var}(X)$  & 0.485  & 0.068  & 0.437  & 0.484  & 0.531 \tabularnewline
$\text{Var}[X|Y>Q_{0.95}(Y)]/\text{Var}(X)$  & 0.433  & 0.092  & 0.369  & 0.432  & 0.489 \tabularnewline
$\text{Var}[X|Y>Q_{0.99}(Y)]/\text{Var}(X)$  & 0.369  & 0.200  & 0.222  & 0.337  & 0.478 \tabularnewline
\hline 
 & \multicolumn{5}{c}{Extremal-Quantile $n=5000$}\tabularnewline
\hline 
$\text{Var}[X|Y>Q_{0.90}(Y)]/\text{Var}(X)$  & 0.481  & 0.030  & 0.461  & 0.481  & 0.501 \tabularnewline
$\text{Var}[X|Y>Q_{0.95}(Y)]/\text{Var}(X)$  & 0.428  & 0.040  & 0.400  & 0.427  & 0.455 \tabularnewline
$\text{Var}[X|Y>Q_{0.99}(Y)]/\text{Var}(X)$  & 0.361  & 0.085  & 0.303  & 0.353  & 0.416 \tabularnewline
\hline 
\end{tabular}
\end{table}

\begin{table}[htbp]
\label{Tab:2} \centering \caption{DGP4M Variance ratios in small samples}
\begin{tabular}{c|ccccc}
\hline 
 & Mean  & Std  & 25\%  & 50\%  & 75\% \tabularnewline
\hline 
 & \multicolumn{5}{c}{Tail-Index $n=1000$}\tabularnewline
\hline 
$\text{Var}_{\text{4M}}[X|Y>Q_{0.90}(Y)]/\text{Var}_{4\textrm{M}}(X)$  & 0.148  & 0.029  & 0.127  & 0.145  & 0.165 \tabularnewline
$\text{Var}_{\text{4M}}[X|Y>Q_{0.95}(Y)]/\text{\ensuremath{\text{Var}_{4\textrm{M}}}}(X)$  & 0.092  & 0.024  & 0.075  & 0.090  & 0.106 \tabularnewline
$\text{Var}_{\text{4M}}[X|Y>Q_{0.99}(Y)]/\text{\ensuremath{\text{Var}_{4\textrm{M}}}}(X)$  & $-$  & $-$  & $-$  & $-$  & $-$ \tabularnewline
\hline 
 & \multicolumn{5}{c}{Tail-Index $n=5000$}\tabularnewline
\hline 
$\text{Var}_{\text{4M}}[X|Y>Q_{0.90}(Y)]/\text{\ensuremath{\text{Var}_{4\textrm{M}}}}(X)$  & 0.148  & 0.013  & 0.139  & 0.148  & 0.156 \tabularnewline
$\text{Var}_{\text{4M}}[X|Y>Q_{0.95}(Y)]/\text{Var}_{4\textrm{M}}(X)$  & 0.090  & 0.010  & 0.083  & 0.090  & 0.096 \tabularnewline
$\text{Var}_{\text{4M}}[X|Y>Q_{0.99}(Y)]/\text{\ensuremath{\text{Var}_{4\textrm{M}}}}(X)$  & 0.045  & 0.011  & 0.038  & 0.044  & 0.051 \tabularnewline
\hline 
 & \multicolumn{5}{c}{Extremal-Quantile $n=1000$}\tabularnewline
\hline 
$\text{Var}_{\text{4M}}[X|Y>Q_{0.90}(Y)]/\text{\ensuremath{\text{Var}_{4\textrm{M}}}}(X)$  & 0.284  & 0.038  & 0.256  & 0.281  & 0.308 \tabularnewline
$\text{Var}_{\text{4M}}[X|Y>Q_{0.95}(Y)]/\text{\ensuremath{\text{Var}_{4\textrm{M}}}}(X)$  & 0.248  & 0.050  & 0.213  & 0.243  & 0.282 \tabularnewline
$\text{Var}_{\text{4M}}[X|Y>Q_{0.99}(Y)]/\text{Var}_{4\textrm{M}}(X)$  & $-$  & $-$  & $-$  & $-$  & $-$ \tabularnewline
\hline 
 & \multicolumn{5}{c}{Extremal-Quantile $n=5000$}\tabularnewline
\hline 
$\text{Var}_{\text{4M}}[X|Y>Q_{0.90}(Y)]/\text{Var}_{4\textrm{M}}(X)$  & 0.287  & 0.017  & 0.275  & 0.287  & 0.298 \tabularnewline
$\text{Var}_{\text{4M}}[X|Y>Q_{0.95}(Y)]/\text{\ensuremath{\text{Var}_{4\textrm{M}}}}(X)$  & 0.251  & 0.022  & 0.236  & 0.251  & 0.266 \tabularnewline
$\text{Var}_{\text{4M}}[X|Y>Q_{0.99}(Y)]/\text{\ensuremath{\text{Var}_{4\textrm{M}}}}(X)$  & 0.213  & 0.047  & 0.181  & 0.209  & 0.241 \tabularnewline
\hline 
\end{tabular}
\end{table}

\begin{table}[htbp]
\label{Tab:3} \centering \caption{DGP2X1M Minimum eigenvalue ratios in small samples}
\begin{tabular}{c|ccccc}
\hline 
 & Mean  & Std  & 25\%  & 50\%  & 75\% \tabularnewline
\hline 
 & \multicolumn{5}{c}{Tail-Index $n=1000$}\tabularnewline
\hline 
$\rho_{\min}(\widehat{\mathbb{E}}[ZZ'|Y>Q_{0.90}(Y)])/\rho_{\min}(\widehat{\mathbb{E}}[ZZ'])$  & 0.949  & 0.110  & 0.869  & 0.951  & 1.020 \tabularnewline
$\rho_{\min}(\widehat{\mathbb{E}}[ZZ'|Y>Q_{0.95}(Y)])/\rho_{\min}(\widehat{\mathbb{E}}[ZZ'])$  & 0.786  & 0.148  & 0.685  & 0.784  & 0.884 \tabularnewline
$\rho_{\min}(\widehat{\mathbb{E}}[ZZ'|Y>Q_{0.99}(Y)])/\rho_{\min}(\widehat{\mathbb{E}}[ZZ'])$  & 0.300  & 0.190  & 0.157  & 0.254  & 0.404 \tabularnewline
\hline 
 & \multicolumn{5}{c}{Tail-Index $n=5000$}\tabularnewline
\hline 
$\rho_{\min}(\widehat{\mathbb{E}}[ZZ'|Y>Q_{0.90}(Y)])/\rho_{\min}(\widehat{\mathbb{E}}[ZZ'])$  & 1.002  & 0.055  & 0.962  & 1.003  & 1.037 \tabularnewline
$\rho_{\min}(\widehat{\mathbb{E}}[ZZ'|Y>Q_{0.95}(Y)])/\rho_{\min}(\widehat{\mathbb{E}}[ZZ'])$  & 0.888  & 0.077  & 0.833  & 0.890  & 0.937 \tabularnewline
$\rho_{\min}(\widehat{\mathbb{E}}[ZZ'|Y>Q_{0.99}(Y)])/\rho_{\min}(\widehat{\mathbb{E}}[ZZ'])$  & 0.470  & 0.118  & 0.386  & 0.464  & 0.547 \tabularnewline
\hline 
 & \multicolumn{5}{c}{Extremal-Quantile $n=1000$}\tabularnewline
\hline 
$\rho_{\min}(\widehat{\mathbb{E}}[ZZ'|Y>Q_{0.90}(Y)])/\rho_{\min}(\widehat{\mathbb{E}}[ZZ'])$  & 0.827  & 0.100  & 0.759  & 0.826  & 0.892 \tabularnewline
$\rho_{\min}(\widehat{\mathbb{E}}[ZZ'|Y>Q_{0.95}(Y)])/\rho_{\min}(\widehat{\mathbb{E}}[ZZ'])$  & 0.786  & 0.130  & 0.690  & 0.782  & 0.874 \tabularnewline
$\rho_{\min}(\widehat{\mathbb{E}}[ZZ'|Y>Q_{0.99}(Y)])/\rho_{\min}(\widehat{\mathbb{E}}[ZZ'])$  & 0.533  & 0.228  & 0.364  & 0.515  & 0.668 \tabularnewline
\hline 
 & \multicolumn{5}{c}{Extremal-Quantile $n=5000$}\tabularnewline
\hline 
$\rho_{\min}(\widehat{\mathbb{E}}[ZZ'|Y>Q_{0.90}(Y)])/\rho_{\min}(\widehat{\mathbb{E}}[ZZ'])$  & 0.846  & 0.045  & 0.814  & 0.846  & 0.875 \tabularnewline
$\rho_{\min}(\widehat{\mathbb{E}}[ZZ'|Y>Q_{0.95}(Y)])/\rho_{\min}(\widehat{\mathbb{E}}[ZZ'])$  & 0.828  & 0.063  & 0.783  & 0.825  & 0.870 \tabularnewline
$\rho_{\min}(\widehat{\mathbb{E}}[ZZ'|Y>Q_{0.99}(Y)])/\rho_{\min}(\widehat{\mathbb{E}}[ZZ'])$  & 0.767  & 0.130  & 0.676  & 0.765  & 0.847 \tabularnewline
\hline 
\end{tabular}
\end{table}

We now turn to the finite-sample results. We do not report the results
for DGP4M at $\tau=0.99$ when $n=1000$, since this corresponds to
only 10 observations in the tail. In such cases, $\mathrm{Var}_{4\mathrm{M}}$
may return NaN when fewer than two observations fall within an interval.
Tables 1--3 show that sampling variability increases at the deepest
thresholds, particularly when $n=1000$. For DGP2X1M, the minimum-eigenvalue
ratios at $\tau=0.99$ are noticeably below their close-to-population
counterparts when $n=1000$, but move toward them as the sample size
increases to $n=5000$. Thus, the finite-sample results keep the qualitative
difference between the two frameworks, but the deepest thresholds
should be read with caution.

\section{An Empirical Illustration}

\label{sec:Empirical-Illustration}

Stock returns are a natural setting for extreme-value methods. Both
tail index regressions and extremal quantile regression have been
applied in this context; see, for example, \citet{ChernozhukovHandbook},
\citet{deHaan03072021}, and \citet{NICOLAUJoE2023}, and the references
therein.

We study daily returns of the S\&P 500 index from January 1, 1929
to December 31, 2024, giving a total of $T=24{,}114$ observations.
The data are downloaded from \textbf{\textcolor{blue}{yahoo!}}finance.
The daily return $Y$ is defined as the log difference of the index
between two consecutive trading days. We examine whether the data
exhibit the degeneracy pattern implied by a varying tail index. The
existing literature offers mixed evidence. For example, \citet{EinmahlEtal2014}
find that the tail index varies over time, whereas \citet{deHaan03072021}
cannot reject the hypothesis of a constant tail index for daily S\&P
500 returns over the period 1988--2012.

We examine this issue by studying the density and variance of $t$
conditional on extreme negative returns. We normalize time $t$ to
the interval $\left(0,1\right]$; for instance, the first observation
corresponds to $t=1/24114$ and the last to $t=1$. As shown in the
simulation studies, the analysis is complicated by the possibility
of multiple modes. We therefore treat major financial crises in the
United States stock market as potential modes of the conditional density
when calculating conditional variances.

We consider two mode specifications. The first specification has four
modes: the 1929 Wall Street Crash ($t=0$), the 1987 Black Monday
crash ($t=0.61$), the 2008 Global Financial Crisis ($t=0.83$), and
the 2020 COVID-19 crash ($t=0.958$). The second specification adds
two more modes, giving six modes in total: the 1973 oil shock and
stagflation crisis ($t=0.46$) and the 2000 dot-com crash ($t=0.75$).

We consider conditional sets $\{Y<Q_{\tau}(Y)\}$ with $\tau=0.1,0.05,0.01,$
and $0.005$. The conditional variance is defined as 
\begin{align*}
\textrm{Var}_{K\textrm{M}}\left(t\mid Y<Q_{\tau}(Y)\right) & =\textrm{Var}\left(t\mid Y<Q_{\tau}(Y),\,t<c_{1}\right)+\sum_{l=1}^{K-2}\textrm{Var}\left(t\mid Y<Q_{\tau}(Y),\,c_{l}\le t<c_{l+1}\right)\\
 & \quad+\textrm{Var}\left(t\mid Y<Q_{\tau}(Y),\,t\ge c_{K-1}\right),
\end{align*}
where $K=4,6$, and $c_{l}$ denotes the midpoint between two adjacent
modes. For example, $c_{1}=(0+0.61)/2$ in the four-mode setting and
$c_{1}=(0+0.46)/2$ in the six-mode setting, with the remaining cutoffs
defined in the same way.

The conditional density is estimated using a histogram estimator.
Define bins 
\[
\mathbb{B}_{j}=\left(\left(j-1\right)h,jh\right],\quad j=1,2,\ldots,1/h,\quad\textrm{with }h=0.01,
\]
and 
\[
\hat{f}_{h}\left(t\mid Y<Q_{\tau}(Y)\right)=\frac{1}{T\tau h}\sum_{t:y_{t}\le Q_{\tau}(Y)}1\left(t\in\mathbb{B}_{j}\right),\quad\textrm{for }t\in\mathbb{B}_{j}.
\]

\begin{figure}
\centering\protect\caption{S\&P 500 1929 -- 2024}
\includegraphics[width=1\textwidth]{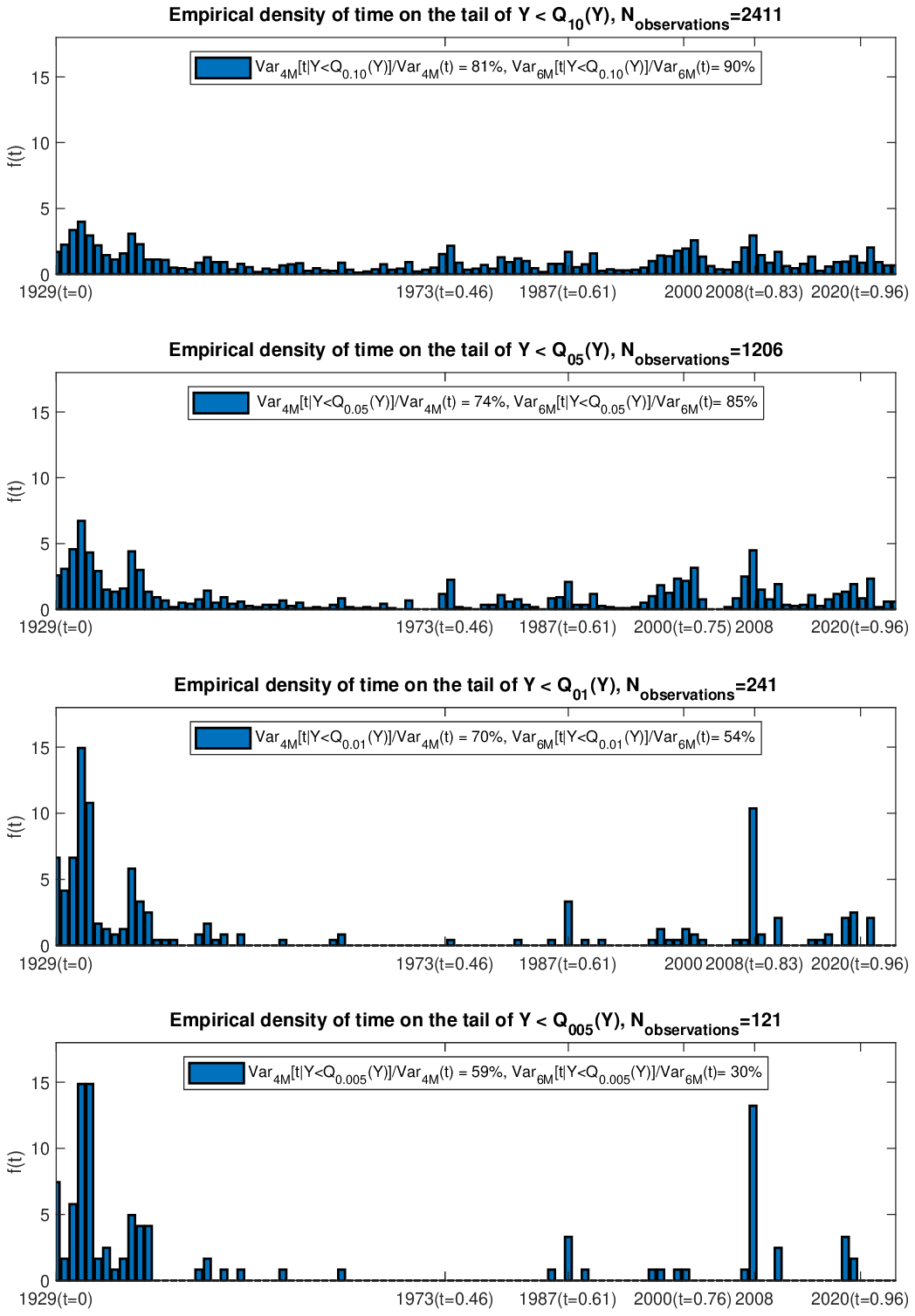} \label{fig:sp500_f1} 
\end{figure}

The estimated densities and conditional variances for both mode specifications
are reported in Figure \ref{fig:sp500_f1}. The effective number of
observations used in the estimation is $T\tau$, which is denoted
as $N_{\textrm{observations}}$ in the figures. As $\tau$ decreases,
the timing of extreme events becomes increasingly concentrated around
the identified modes. This pattern indicates that either $\alpha(t)$,
$k(t)$, or both vary over time.

Examining the conditional variances, we find that the conditional
density does not degenerate as rapidly as in the tail index regression
framework with a varying $\alpha(t)$ observed in the simulation studies.
However, the conditional variance also does not stabilize as in the
extremal quantile regression framework. Consequently, the empirical
evidence is somewhat mixed. Nevertheless, $\textrm{Var}_{4\textrm{M}}\left(t\mid Y<Q_{0.005}(Y)\right)$
and $\textrm{Var}_{6\textrm{M}}\left(t\mid Y<Q_{0.005}(Y)\right)$
remain at 59 percent and 30 percent of the unconditional variance,
respectively, suggesting that the conditional distribution does not
degenerate. Taken together, these results are more consistent with
the extremal quantile regression pattern than with the strong degeneracy
generated by the varying-tail-index designs in our simulations. We
emphasize, however, that this empirical exercise is only illustrative.
The diagnostic does not formally distinguish a varying tail index
from a varying scale function, and the quantitative results remain
sensitive to the choice of crisis modes.

\begin{figure}
\centering\caption{S\&P 500 1988 -- 2012}
\includegraphics[width=1\textwidth]{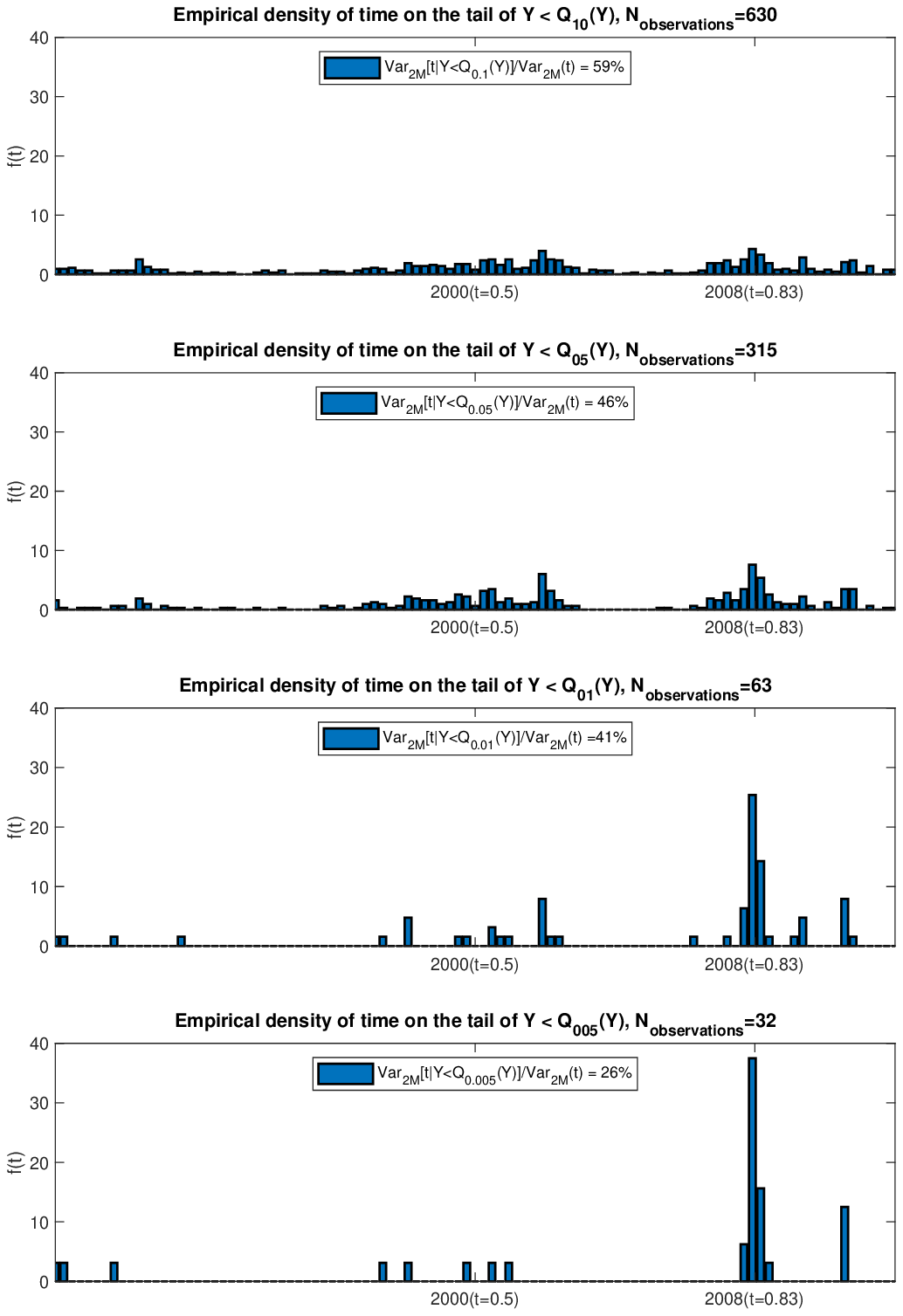} \label{fig:sp500_f2} 
\end{figure}

We next examine the subperiod from January 1, 1988 to December 31,
2012, studied by \citet{EinmahlEtal2014} and \citet{deHaan03072021},
for which $T=6{,}302$. We normalize $t$ to $\left(0,1\right]$ as
before and consider a two-mode specification: the 2000 dot-com crash
($t=0.5$) and the 2008 Global Financial Crisis ($t=0.83$). The conditional
density and the two-mode conditional variance, $\textrm{Var}_{2\textrm{M}}\left(t\mid Y<Q_{\tau}(Y)\right)$,
are computed as before. The results are reported in Figure \ref{fig:sp500_f2}.

The pattern is similar to the one in Figure \ref{fig:sp500_f1}. It
implies that either $\alpha(t)$, $k(t)$, or both vary over time,
while the evidence regarding the more appropriate framework remains
mixed. However, $\textrm{Var}_{2\textrm{M}}\left(t\mid Y<Q_{\tau}(Y)\right)/\textrm{Var}_{2\textrm{M}}(t)$
equals 59 percent, 46 percent, 41 percent, and 26 percent for $\tau=0.1,0.05,0.01,$
and $0.005$, respectively. Given that $N_{\textrm{observations}}$
is only 32 for $\tau=0.005$, the patterns for $\tau=0.1,0.05,$ and
$0.01$ are more consistent with the extremal quantile regression
pattern.

The previous analysis assumes i.i.d. observations, whereas financial
return series may exhibit GARCH-type dynamics. In Appendix \ref{APP:C},
we show that the sample counterpart of $\mathbb{E}\left(XX'|Y>w\right)$
remains consistent at a rate that accommodates a degenerating minimum
eigenvalue under $\alpha$-mixing. This result provides theoretical
support for the Gram-matrix component of the diagnostic under a class
of weakly dependent processes; it does not by itself establish robustness
of the entire empirical diagnostic, which also involves sample quantiles
and the choice of modes.

One limitation of this empirical illustration is that the results
depend on modes that we choose by hand.

\section{Conclusion}

\label{sec:Conclusion}

In this paper, we study the rank condition and the conditional density
of explanatory variables in tail index regressions. These conditions
can fail because the estimation sample is selected in a way that depends
on the explanatory variables. We provide an asymptotic treatment that
accounts for the degenerate Gram matrix for the linear tail index
model. This issue does not arise in the extremal quantile regression
framework because explanatory variables have a weaker effect on tail
behavior than in the tail index regression framework. These different
behaviors provide a simple diagnostic for checking whether the data
show the degeneracy pattern of a varying tail index. The diagnostic
is illustrative rather than a formal model-selection procedure.

\appendix
\begin{center}
{\LARGE\textbf{{}{}{}{}{}Appendix}}{\LARGE{}{}{}{} }{\LARGE\par}
\par\end{center}

\begin{center}
 
\par\end{center}

\section{Lemmas and Proofs}

\label{APP:A}

\begin{lemma}\label{LE:1}The results in Theorem \ref{TH:1} imply
(\ref{eq:var->0}).

\end{lemma}

\noindent\textbf{Proof.} Note that for $z>\underline{u}>0$, $\log z-\log\underline{u}=\frac{z-\underline{u}}{\xi}\leq\frac{z-\underline{u}}{\underline{u}},$
for some $\xi\in\left(\underline{u},z\right)$ by the mean value theorem.
Therefore, 
\[
\mathbb{E}(X_{1}\mid Y>w)-\log\left(\underline{u}\right)=\mathbb{E}(\log Z\mid Y>w)-\log\left(\underline{u}\right)\leq\left[\mathbb{E}\left(Z\mid Y>w\right)-\underline{u}\right]/\underline{u}\to0.
\]
For the variance, let $Z'$ be an i.i.d. copy. Then 
\begin{align*}
\textrm{Var}\left(X_{1}\mid Y>w\right) & =\frac{1}{2}\mathbb{E}\left[\left(\log Z-\log Z'\right)^{2}\mid Y>w\right]\leq\frac{1}{2\underline{u}^{2}}\mathbb{E}\left[\left(Z-Z'\right)^{2}\mid Y>w\right]\\
 & =\frac{1}{\underline{u}^{2}}\textrm{Var}\left(Z\mid Y>w\right)\to0,
\end{align*}
as desired.\hfill{}$\blacksquare$

\begin{lemma}\label{LE:2}The results in Corollary \ref{Coro:1}
imply (\ref{eq:varX1bound}).

\end{lemma}

\noindent\textbf{Proof.} The second inequality in (\ref{eq:varX1bound})
is directly from the proof of Lemma \ref{LE:1}. Note for $x_{1}<x_{1}'<\log\bar{u}=\bar{u}_{x}<\infty,$
\[
0<\exp\left(x_{1}'\right)-\exp\left(x_{1}\right)=\exp\left(\xi\right)\left(x_{1}'-x_{1}\right)\leq\bar{u}\left(x_{1}'-x_{1}\right),
\]
for some $\xi\in\left(x_{1},x_{1}'\right)$ by the mean value theorem.
Let $X_{1}'$ be an i.i.d. copy. The first inequality is obtained
by 
\begin{align*}
\textrm{Var}\left(Z\mid Y>w\right) & =\frac{1}{2}\mathbb{E}\left[\left(\exp X_{1}-\exp X_{1}'\right)^{2}\mid Y>w\right]\\
 & \leq\frac{\bar{u}^{2}}{2}\mathbb{E}\left[\left(X_{1}-X_{1}'\right)^{2}\mid Y>w\right]=\bar{u}^{2}\textrm{Var}\left(X_{1}\mid Y>w\right),
\end{align*}
as desired.\hfill{}$\blacksquare$

\noindent\textbf{Proof of Theorem \ref{TH:1}.} We first derive the
conditional density of $Z|Y>w.$

Using Bayes' theorem, we have the first result: 
\begin{align}
f\left(z|Y>w\right) & =\frac{\Pr\left(Y>w|z\right)f\left(z\right)}{\int_{\underline{u}}^{\bar{u}}\Pr\left(Y>w|z\right)f\left(z\right)\textrm{d}z}\nonumber \\
 & =\frac{w^{-z}f\left(z\right)}{\int_{\underline{u}}^{\bar{u}}w^{-z}f\left(z\right)\textrm{d}z}=\frac{w^{\underline{u}-z}f\left(z\right)}{\int_{\underline{u}}^{\bar{u}}w^{\underline{u}-z}f\left(z\right)\textrm{d}z}\nonumber \\
 & \leq\frac{\bar{c}}{\underline{c}}\frac{w^{\underline{u}-z}}{\int_{\underline{u}}^{\bar{u}}w^{\underline{u}-z}\textrm{d}z}=\frac{\bar{c}}{\underline{c}}\frac{w^{-\left(z-\underline{u}\right)}\log w}{1-w^{-\left(\bar{u}-\underline{u}\right)}}.\label{eq:f(z|Y>w)}
\end{align}

Using the second line in (\ref{eq:f(z|Y>w)}), 
\begin{align*}
\mathbb{E}\left(Z|Y>w\right) & =\int_{\underline{u}}^{\bar{u}}zf\left(z|Y>w\right)\textrm{d}z\\
 & =\frac{\int_{\underline{u}}^{\bar{u}}w^{\underline{u}-z}zf\left(z\right)\textrm{d}z}{\int_{\underline{u}}^{\bar{u}}w^{\underline{u}-z}f\left(z\right)\textrm{d}z}=\frac{\int_{\underline{u}}^{\underline{u}+\epsilon}w^{\underline{u}-z}zf\left(z\right)\textrm{d}z+\int_{\underline{u}+\epsilon}^{\bar{u}}w^{\underline{u}-z}zf\left(z\right)\textrm{d}z}{\int_{\underline{u}}^{\underline{u}+\epsilon}w^{\underline{u}-z}f\left(z\right)\textrm{d}z+\int_{\underline{u}+\epsilon}^{\bar{u}}w^{\underline{u}-z}f\left(z\right)\textrm{d}z}\\
 & \equiv\frac{D_{1\epsilon}\left(w\right)+R_{1\epsilon}\left(w\right)}{D_{2\epsilon}\left(w\right)+R_{2\epsilon}\left(w\right)},
\end{align*}
for some small $\epsilon>0$. We claim that as $w\rightarrow\infty,$
\[
R_{1\epsilon}\left(w\right)=o\left(D_{2\epsilon}\left(w\right)\right)\textrm{ and }R_{2\epsilon}\left(w\right)=o\left(D_{2\epsilon}\left(w\right)\right).
\]
We prove this claim at the end.

Note that 
\[
\frac{D_{1\epsilon}\left(w\right)}{D_{2\epsilon}\left(w\right)}=\frac{\int_{\underline{u}}^{\underline{u}+\epsilon}w^{\underline{u}-z}zf\left(z\right)\textrm{d}z}{\int_{\underline{u}}^{\underline{u}+\epsilon}w^{\underline{u}-z}f\left(z\right)\textrm{d}z}\leq\frac{\left(\underline{u}+\epsilon\right)\int_{\underline{u}}^{\underline{u}+\epsilon}w^{\underline{u}-z}f\left(z\right)\textrm{d}z}{\int_{\underline{u}}^{\underline{u}+\epsilon}w^{\underline{u}-z}f\left(z\right)\textrm{d}z}=\underline{u}+\epsilon.
\]
Therefore when $w$ is large enough, $\mathbb{E}\left(Z|Y>w\right)\leq\underline{u}+2\epsilon.$
Since $\epsilon$ can be arbitrarily small, we have 
\begin{equation}
\mathbb{E}\left(Z|Y>w\right)\rightarrow\underline{u},\textrm{ as }w\rightarrow\infty.\label{eq:EZ|Y}
\end{equation}
Similarly, 
\begin{equation}
\mathbb{E}\left(Z^{2}|Y>w\right)\rightarrow\underline{u}^{2},\textrm{ as }w\rightarrow\infty.\label{eq:EZ^2|Y}
\end{equation}
By (\ref{eq:EZ|Y}) and (\ref{eq:EZ^2|Y}), we must have $\textrm{Var}\left(Z|Y>w\right)=\mathbb{E}\left(Z^{2}|Y>w\right)-\left[\mathbb{E}\left(Z|Y>w\right)\right]^{2}\rightarrow0$
$\textrm{as }w\rightarrow\infty.$

We now show the claim. Note that Assumption \ref{A:1} implies 
\begin{equation}
D_{2\epsilon}\left(w\right)\geq\int_{\underline{u}}^{\underline{u}+\epsilon/2}w^{\underline{u}-z}f\left(z\right)\textrm{d}z\geq\frac{\epsilon}{2}w^{-\epsilon/2}\underline{c}.\label{eq:D2w}
\end{equation}
On the other hand, Assumption \ref{A:1} guarantees 
\begin{equation}
\max\left\{ R_{1\epsilon}\left(w\right),R_{2\epsilon}\left(w\right)\right\} \leq\left(\bar{u}-\underline{u}\right)\max\left\{ 1,\bar{u}\right\} w^{-\epsilon}\bar{c}.\label{eq:maxR}
\end{equation}
Since $\epsilon w^{-\epsilon/2}\gg w^{-\epsilon}$ as $w\rightarrow\infty$
for a fixed $\epsilon,$ (\ref{eq:D2w}) and (\ref{eq:maxR}) imply
that $D_{2\epsilon}\left(w\right)\gg\max\left\{ R_{1\epsilon}\left(w\right),R_{2\epsilon}\left(w\right)\right\} $
as desired. \hfill{}$\blacksquare$

\noindent\textbf{Proof of Corollary \ref{Coro:1}.} We first derive
the conditional density of $Z:$ 
\begin{align*}
f\left(z|Y>w\right) & =\frac{\Pr\left(Y>w|z\right)f\left(z\right)}{\int_{\underline{u}}^{\bar{u}}\Pr\left(Y>w|z\right)f\left(z\right)\textrm{d}z}\\
 & =\frac{w^{-z}}{\int_{\underline{u}}^{\bar{u}}w^{-z}\textrm{d}z}=\frac{w^{-z}\log w}{w^{-\underline{u}}-w^{-\bar{u}}}=\frac{w^{-\left(z-\underline{u}\right)}\log w}{1-w_{}^{\underline{u}-\bar{u}}},\textrm{ for }z\in\left[\underline{u},\bar{u}\right].
\end{align*}
Based on it, 
\begin{align*}
\mathbb{E}\left(Z|Y>w\right) & =\int_{\underline{u}}^{\bar{u}}z\frac{w^{\underline{u}-z}\log w}{1-w_{}^{\underline{u}-\bar{u}}}\textrm{d}z=\underline{u}+\frac{1}{\log w}-\frac{\left(\bar{u}-\underline{u}\right)w^{\underline{u}-\bar{u}}}{1-w^{\underline{u}-\bar{u}}}=\underline{u}+\frac{1}{\log w}+o\left(\frac{1}{\log w}\right).
\end{align*}
and 
\begin{align*}
\mathbb{E}\left(Z^{2}|Y>w\right) & =\int_{\underline{u}}^{\bar{u}}z^{2}\frac{w^{\underline{u}-z}\log w}{1-w^{\underline{u}-\bar{u}}}\textrm{d}z\\
 & =\frac{\underline{u}^{2}-\bar{u}^{2}w^{\underline{u}-\bar{u}}}{1-w^{\underline{u}-\bar{u}}}+\frac{2}{\log w}\mathbb{E}\left(Z|Y>w\right)\\
 & =\underline{u}^{2}-\frac{\left(\bar{u}^{2}-\underline{u}^{2}\right)w^{\underline{u}-\bar{u}}}{1-w^{\underline{u}-\bar{u}}}+\frac{2}{\log w}\mathbb{E}\left(Z|Y>w\right).
\end{align*}
We proceed to calculate its variance, 
\begin{align*}
\textrm{Var}\left(Z|Y>w\right) & =\mathbb{E}\left(Z^{2}|Y>w\right)-\left[\mathbb{E}\left(Z|Y>w\right)\right]^{2}\\
 & =\frac{1}{\left(\log w\right)^{2}}-\frac{\left(\bar{u}-\underline{u}\right)^{2}w^{-\left(\bar{u}-\underline{u}\right)}}{\left(1-w^{-\left(\bar{u}-\underline{u}\right)}\right)^{2}}\\
 & =\frac{1}{\left(\log w\right)^{2}}+o\left(\frac{1}{\left(\log w\right)^{2}}\right)\text{,}
\end{align*}
as desired. As a side note, the variance above can be shown to be
positive for any $w$.\hfill{}$\blacksquare$

\noindent\textbf{Proof of Corollary \ref{Coro:2}.} The proof proceeds
in three steps. Step 1 relaxes the lower-bound requirement in Assumption
\ref{A:1}. Step 2 derives the behavior of the density of $\tilde{X}_{1}$
near the lower endpoint of its support. Step 3 combines the two.

\noindent\textbf{Step 1. A weaker lower bound in Theorem \ref{TH:1}.}
Suppose that $Z$ has support $\left[\underline{u},\bar{u}\right]$
with density $f$ satisfying $f(z)\leq\bar{c}<\infty$ for all $z$
and, for some constants $k\geq0$, $\delta\in\left(0,\bar{u}-\underline{u}\right]$
and $\underline{c}>0$, 
\[
f(z)\geq\underline{c}\left(z-\underline{u}\right)^{k},\qquad z\in\left[\underline{u},\underline{u}+\delta\right].
\]
We claim that the conclusions of Theorem \ref{TH:1} remain valid,
with the density bound replaced by 
\[
f(z\mid Y>w)\leq Cw^{-\left(z-\underline{u}\right)}\left(\log w\right)^{k+1},\qquad z\in\left[\underline{u},\bar{u}\right],
\]
for some $C>0$.

Only the lower bound (\ref{eq:D2w}) on $D_{2\epsilon}\left(w\right)$
in the proof of Theorem \ref{TH:1} has to be modified. Fix $\epsilon\leq2\delta$.
Substituting $t=\left(z-\underline{u}\right)\log w$, 
\[
\int_{\underline{u}}^{\underline{u}+\epsilon/2}w^{\underline{u}-z}f\left(z\right)\textrm{d}z\geq\frac{\underline{c}}{\left(\log w\right)^{k+1}}\int_{0}^{\left(\epsilon/2\right)\log w}t^{k}e^{-t}\textrm{d}t\geq\frac{C}{\left(\log w\right)^{k+1}}
\]
for $w$ large enough, because the last integral converges to $\Gamma(k+1)$
as $w\to\infty$. The bound (\ref{eq:maxR}) on $\max\left\{ R_{1\epsilon}\left(w\right),R_{2\epsilon}\left(w\right)\right\} $
only uses $f\leq\bar{c}$ and is therefore unchanged. Since $\left(\log w\right)^{-(k+1)}\gg w^{-\epsilon}$
as $w\to\infty$ for a fixed $\epsilon$, we still have $D_{2\epsilon}\left(w\right)\gg\max\left\{ R_{1\epsilon}\left(w\right),R_{2\epsilon}\left(w\right)\right\} $,
and the remaining arguments in the proof of Theorem \ref{TH:1} apply
without change. Hence $\mathbb{E}(Z\mid Y>w)\to\underline{u}$ and
$\operatorname{Var}(Z\mid Y>w)\to0$. For the density bound, use the
second line of (\ref{eq:f(z|Y>w)}): the numerator is at most $\bar{c}w^{-\left(z-\underline{u}\right)}$,
while, taking $\epsilon=2\delta$ in the last display, the denominator
satisfies $\int_{\underline{u}}^{\bar{u}}w^{\underline{u}-z}f\left(z\right)\textrm{d}z\geq C\left(\log w\right)^{-(k+1)}$.

\noindent\textbf{Step 2. The density of $\tilde{X}_{1}$ near $\underline{\tilde{u}}_{x_{1}}$.}
Let $f_{j}$ denote the density of $X_{j}$, supported on $\left[\underline{u}_{x_{j}},\bar{u}_{x_{j}}\right]$,
so that $\underline{\tilde{u}}_{x_{1}}=\sum_{j=1}^{p}\underline{u}_{x_{j}}$,
and let $\bar{h}=\min_{j}\left(\bar{u}_{x_{j}}-\underline{u}_{x_{j}}\right)>0$.
By Assumption \ref{A:1}, $0<\underline{c}_{x}\leq f_{j}\leq\bar{c}_{x}<\infty$
on $\left[\underline{u}_{x_{j}},\bar{u}_{x_{j}}\right]$ for each
$j$. For $0<h\leq\bar{h}$, independence gives 
\[
f\left(\underline{\tilde{u}}_{x_{1}}+h\right)=\int_{\Delta(h)}\prod_{j=1}^{p}f_{j}\!\left(\underline{u}_{x_{j}}+t_{j}\right)\textrm{d}t_{1}\cdots\textrm{d}t_{p-1},\qquad t_{p}=h-\sum_{j=1}^{p-1}t_{j},
\]
where $\Delta(h)=\{(t_{1},\ldots,t_{p-1}):t_{j}\geq0,\ \sum_{j=1}^{p-1}t_{j}\leq h\}$
has volume $h^{p-1}/(p-1)!$. On $\Delta(h)$ we have $0\leq t_{j}\leq h\leq\bar{h}$
for every $j$, so each $\underline{u}_{x_{j}}+t_{j}$ lies in the
support of $X_{j}$ and the integrand is bounded between $\underline{c}_{x}^{p}$
and $\bar{c}_{x}^{p}$. Therefore 
\[
\frac{\underline{c}_{x}^{p}}{(p-1)!}h^{p-1}\leq f\left(\underline{\tilde{u}}_{x_{1}}+h\right)\leq\frac{\bar{c}_{x}^{p}}{(p-1)!}h^{p-1},\qquad0<h\leq\bar{h}.
\]
In particular, for $p\geq2$ the density of $\tilde{X}_{1}$ vanishes
at $\underline{\tilde{u}}_{x_{1}}$, so Assumption \ref{A:1} does
not hold for $\tilde{X}_{1}$. The density of $\tilde{X}_{1}$ is
nonetheless bounded above on its whole support, since it is a convolution
of bounded densities.

\noindent\textbf{Step 3. Conclusion.} Let $\tilde{Z}=\alpha(X)=\exp(\tilde{X}_{1})$,
with support $\left[\underline{\tilde{u}},\bar{\tilde{u}}\right]$
where $\underline{\tilde{u}}=\exp(\underline{\tilde{u}}_{x_{1}})>0$
and $\bar{\tilde{u}}=\exp(\sup\tilde{X}_{1})<\infty$. By the change
of variables, $f_{\tilde{Z}}(z)=f\left(\log z\right)/z$, so $f_{\tilde{Z}}$
is bounded above by Step 2. By the mean value theorem, for $z\in\left[\underline{\tilde{u}},\bar{\tilde{u}}\right]$,
\[
\frac{z-\underline{\tilde{u}}}{\bar{\tilde{u}}}\leq\log z-\underline{\tilde{u}}_{x_{1}}\leq\frac{z-\underline{\tilde{u}}}{\underline{\tilde{u}}}.
\]
Choose $\delta>0$ small enough that $\delta/\underline{\tilde{u}}\leq\bar{h}$.
Then, for $z\in\left[\underline{\tilde{u}},\underline{\tilde{u}}+\delta\right]$,
applying Step 2 with $h=\log z-\underline{\tilde{u}}_{x_{1}}\in(0,\bar{h}]$
gives 
\[
f_{\tilde{Z}}(z)=\frac{f\left(\log z\right)}{z}\geq\frac{\underline{c}_{x}^{p}}{(p-1)!\,\bar{\tilde{u}}^{p}}\left(z-\underline{\tilde{u}}\right)^{p-1}.
\]
Hence $\tilde{Z}$ satisfies the condition of Step 1 with $k=p-1$,
and Step 1 yields 
\[
\mathbb{E}(\tilde{Z}\mid Y>w)\to\underline{\tilde{u}}\quad\textrm{and}\quad\operatorname{Var}(\tilde{Z}\mid Y>w)\to0.
\]
Since $\tilde{X}_{1}=\log\tilde{Z}$ and $\underline{\tilde{u}}>0$,
Lemma \ref{LE:1} applies with $\left(Z,X_{1}\right)$ replaced by
$(\tilde{Z},\tilde{X}_{1})$, which gives $\mathbb{E}(\tilde{X}_{1}\mid Y>w)\to\underline{\tilde{u}}_{x_{1}}$
and $\operatorname{Var}(\tilde{X}_{1}\mid Y>w)\to0$, as desired.
\hfill{}$\blacksquare$

\noindent\textbf{Proof of Corollary \ref{Coro:2-1}.} \textbf{Part
1.} By Bayes' rule, 
\[
f(z\mid Y>w)=\frac{\Pr(Y>w\mid Z=z)f(z)}{\Pr(Y>w)}=\frac{w^{-z}f(z)}{\int_{\underline{u}}^{\infty}w^{-z}f(z)\textrm{d}z}=\frac{w^{-(z-\underline{u})}f(z)}{\int_{\underline{u}}^{\infty}w^{-(z-\underline{u})}f(z)\textrm{d}z}.
\]

\noindent By the assumption in the Corollary, the denominator satisfies
\begin{equation}
\int_{\underline{u}}^{\infty}w^{-(z-\underline{u})}f(z)\textrm{d}z\ge\underline{c}\int_{\underline{u}}^{\underline{u}+\delta}w^{-(z-\underline{u})}\textrm{d}z=\frac{\underline{c}\left(1-w^{-\delta}\right)}{\log w}.\label{eq:denominator}
\end{equation}

\noindent Substituting this result into the expression for $f(z\mid Y>w)$,
for any fixed $z>\underline{u}$, 
\[
f(z\mid Y>w)\le\frac{w^{-(z-\underline{u})}f(z)}{\underline{c}(1-w^{-\delta})/\log w}=\frac{\bar{c}}{\underline{c}}\frac{w^{-(z-\underline{u})}\log w}{1-w^{-\delta}}.
\]
\textbf{Part 2.1. On the first conditional moment.} For any fixed
small $\varepsilon>0$, 
\[
\mathbb{E}(Z-\underline{u}\mid Y>w)=\mathbb{E}\big[(Z-\underline{u})1\{Z\le\underline{u}+\varepsilon\}\mid Y>w\big]+\mathbb{E}\big[(Z-\underline{u})1\{Z>\underline{u}+\varepsilon\}\mid Y>w\big].
\]
For the first term, 
\[
\mathbb{E}\big[(Z-\underline{u})1\{Z\le\underline{u}+\varepsilon\}\mid Y>w\big]\leq\varepsilon.
\]
For the second term, using Hölder's inequality, 
\begin{align*}
\mathbb{E}\big[(Z-\underline{u})1\{Z>\underline{u}+\varepsilon\}\mid Y>w\big] & \le\Big(\mathbb{E}[(Z-\underline{u})^{1+\eta}\mid Y>w]\Big)^{1/(1+\eta)}\left[\Pr(Z>\underline{u}+\varepsilon\mid Y>w)\right]^{\eta/(1+\eta)}\\
 & \le C\cdot\left[\Pr(Z>\underline{u}+\varepsilon\mid Y>w)\right]{}^{\eta/(1+\eta)},
\end{align*}
for some positive $C$ due to the moment conditions in the Corollary.
Note that

\begin{align*}
\Pr(Z>\underline{u}+\varepsilon\mid Y>w) & =\frac{\int_{\underline{u}+\varepsilon}^{\infty}w^{-(z-\underline{u})}f(z)\,\textrm{d}z}{\int_{\underline{u}}^{\infty}w^{-(z-\underline{u})}f(z)\,\textrm{d}z}\leq\frac{w^{-\varepsilon}\log w}{\underline{c}(1-w^{-\delta})}\rightarrow0\textrm{ for any fixed }\varepsilon\textrm{ as }w\rightarrow\infty,
\end{align*}
where the inequality holds by $\int_{\underline{u}+\varepsilon}^{\infty}w^{-(z-\underline{u})}f(z)\textrm{d}z\leq w^{-\varepsilon}\int_{\underline{u}+\varepsilon}^{\infty}f(z)\textrm{d}z\leq w^{-\varepsilon}$
and the result in (\ref{eq:denominator}) for the denominator.

Hence, for any $\varepsilon>0$, 
\[
0\le\limsup_{w\to\infty}\left[\mathbb{E}(Z\mid Y>w)-\underline{u}\right]\le\varepsilon.
\]
Since $\varepsilon$ is arbitrary, 
\[
\mathbb{E}(Z\mid Y>w)\to\underline{u}.
\]

\noindent\textbf{Part 2.2. On the conditional variance. }Similar
to Part 2.1, we can show that 
\begin{equation}
\mathbb{E}\left[\left.\left(Z-\underline{u}\right)^{2}\right|Y>w\right]\rightarrow0\label{eq:(z-u_)^2|Y>2}
\end{equation}
by decomposing 
\[
\mathbb{E}\left[\left.\left(Z-\underline{u}\right)^{2}\right|Y>w\right]=\mathbb{E}\left[\left.\left(Z-\underline{u}\right)^{2}1\{Z\le\underline{u}+\varepsilon\}\right|Y>w\right]+\mathbb{E}\left[\left.\left(Z-\underline{u}\right)^{2}1\{Z>\underline{u}+\varepsilon\}\right|Y>w\right],
\]
and using $\sup_{w>0}\mathbb{E}\left[\left.\left(Z-\underline{u}\right)^{2+\eta}\right|Y>w\right]<\infty.$

Then 
\begin{align*}
\textrm{Var}\left(Z|Y>w\right) & =\mathbb{E}\left[\left.Z^{2}\right|Y>w\right]-\left[\mathbb{E}\left(\left.Z\right|Y>w\right)\right]^{2}\\
 & =\mathbb{E}\left[\left.\left(Z-\underline{u}\right)^{2}\right|Y>w\right]+2\underline{u}\mathbb{E}\left(\left.Z\right|Y>w\right)-\underline{u}^{2}-\left[\mathbb{E}\left(\left.Z\right|Y>w\right)\right]^{2}\\
 & \rightarrow0,
\end{align*}
using the result in Part 2.1 and (\ref{eq:(z-u_)^2|Y>2}). \hfill{}$\blacksquare$

\begin{lemma}\label{LE:3}Let $A$ and $B$ be $p\times p$ symmetric
positive semidefinite matrices, and let $A^{1/2}$ and $B^{1/2}$
denote their symmetric positive semidefinite square roots. If $\rho_{\min}(A)+\rho_{\min}(B)>0$,
then 
\[
\left\Vert A^{1/2}-B^{1/2}\right\Vert _{F}\leq\frac{\left\Vert A-B\right\Vert _{F}}{\rho_{\min}(A)^{1/2}+\rho_{\min}(B)^{1/2}},
\]
where $\left\Vert \cdot\right\Vert _{F}$ denotes the Frobenius norm.

\end{lemma}

\noindent\textbf{Proof.} Write $X=A^{1/2}=U\Lambda U'$ and $Y=B^{1/2}=VMV'$
for spectral decompositions with $U,V$ orthogonal, $\Lambda=\textrm{diag}(\lambda_{1},\ldots,\lambda_{p})$
and $M=\textrm{diag}(\mu_{1},\ldots,\mu_{p})$ nonnegative. Let $D=X-Y$.
Then 
\[
A-B=X^{2}-Y^{2}=X\left(X-Y\right)+\left(X-Y\right)Y=XD+DY,
\]
so that $U'(A-B)V=\Lambda\left(U'DV\right)+\left(U'DV\right)M$, that
is, element by element, 
\[
\left[U'(A-B)V\right]_{jk}=\left(\lambda_{j}+\mu_{k}\right)\left[U'DV\right]_{jk},\qquad1\leq j,k\leq p.
\]
Since $\lambda_{j}\geq\rho_{\min}(A)^{1/2}$ and $\mu_{k}\geq\rho_{\min}(B)^{1/2}$
for all $j,k$, it follows that 
\[
\left|\left[U'DV\right]_{jk}\right|\leq\frac{\left|\left[U'(A-B)V\right]_{jk}\right|}{\rho_{\min}(A)^{1/2}+\rho_{\min}(B)^{1/2}}.
\]
Squaring, summing over $j$ and $k$, and using the orthogonal invariance
of the Frobenius norm gives the result. See also \citet[Chapter X]{Bhatia1997}
for related perturbation bounds. \hfill{}$\blacksquare$

\noindent\textbf{Proof of Theorem \ref{TH:2}.} We only show the
part that differs from the proof in \citet{WangTsai2009}, namely
how to accommodate a population Gram matrix whose eigenvalues may
have different rates. The rest is the same as in \citet{WangTsai2009}.

Let $\gamma=\theta-\theta^{*}$ and 
\[
\mathcal{K}_{n}^{*}\left(\gamma\right)=\sum_{i=1}^{n}\left\{ \exp\left[x_{i}'\left(\gamma+\theta^{*}\right)\right]\log\left(\left.y_{i}\right/w_{n}\right)-x_{i}'\left(\gamma+\theta^{*}\right)\right\} I\left(y_{i}>w_{n}\right).
\]
Recall that $\underline{\lambda}_{n}=\rho_{\min}(\bar{\varSigma}_{w_{n}})$.
For a fixed $(p+1)\times1$ vector $u$, define the matrix-normalized
local parameter 
\[
\gamma_{n}(u)=\left[n\Pr(Y>w_{n})\right]^{-1/2}\bar{\varSigma}_{w_{n}}^{-1/2}u,\qquad\text{or equivalently}\qquad u=\sqrt{n\Pr(Y>w_{n})}\,\bar{\varSigma}_{w_{n}}^{1/2}\gamma.
\]
The second-order Taylor expansion around zero gives, uniformly over
bounded $u$, 
\begin{align}
\mathcal{K}_{n}^{*}\left(\gamma_{n}(u)\right)-\mathcal{K}_{n}^{*}(0)= & \;u'\left[\left[n\Pr(Y>w_{n})\right]^{-1/2}\bar{\varSigma}_{w_{n}}^{-1/2}\dot{\mathcal{K}}_{n}^{*}(0)\right]\nonumber \\
 & +\frac{1}{2}u'\left\{ \bar{\varSigma}_{w_{n}}^{-1/2}\left[\left[n\Pr(Y>w_{n})\right]^{-1}\ddot{\mathcal{K}}_{n}^{*}(0)\right]\bar{\varSigma}_{w_{n}}^{-1/2}\right\} u+o_{P}(1).\label{eq:Taylor_expansion}
\end{align}
To verify that the remainder in (\ref{eq:Taylor_expansion}) is negligible
uniformly over $\|u\|\leq K$ for any fixed $K<\infty$, let $v_{i}=\bar{\varSigma}_{w_{n}}^{-1/2}x_{i}$
and $\xi_{ni}(u)=x_{i}'\gamma_{n}(u)=\left[n\Pr(Y>w_{n})\right]^{-1/2}u'v_{i}$,
and let $\mathcal{R}_{n}(u)$ denote the exact remainder, that is,
the difference between $\mathcal{K}_{n}^{*}(\gamma_{n}(u))-\mathcal{K}_{n}^{*}(0)$
and the first two terms on the right-hand side of (\ref{eq:Taylor_expansion}).
Since $\exp\left(x_{i}'\theta^{*}\right)\log\left(\left.y_{i}\right/w_{n}\right)=\epsilon_{ni}+1$,
where $\epsilon_{ni}$ is defined in (\ref{eq:epsilon_ni}) below,
a third-order Taylor expansion of the exponential term gives 
\[
\mathcal{R}_{n}(u)=\sum_{i=1}^{n}\left(\epsilon_{ni}+1\right)r\left(\xi_{ni}(u)\right)I(y_{i}>w_{n}),\qquad\left|r(\xi)\right|\leq\left|\xi\right|^{3}e^{\left|\xi\right|}/6.
\]
The moment condition implies $\mathbb{E}\left(\left.\|v_{i}\|^{4}\right|y_{i}>w_{n}\right)\leq C\underline{\lambda}_{n}^{-2}$,
so a union bound and (\ref{eq:varying_rank}) give 
\[
\max_{i:y_{i}>w_{n}}\frac{\|v_{i}\|}{\sqrt{n\Pr(Y>w_{n})}}=o_{P}(1).
\]
Hence $\sup_{\|u\|\leq K}\max_{i:y_{i}>w_{n}}\left|\xi_{ni}(u)\right|=o_{P}(1)$,
so the exponential factor in $r(\cdot)$ is bounded and, with probability
approaching one, 
\[
\sup_{\|u\|\leq K}\left|\mathcal{R}_{n}(u)\right|\leq C\frac{\sum_{i=1}^{n}\left(\epsilon_{ni}+1\right)\|v_{i}\|^{3}I(y_{i}>w_{n})}{\left[n\Pr(Y>w_{n})\right]^{3/2}},
\]
for some $C$ depending only on $K$. Moreover, $\mathbb{E}(v_{i}v_{i}'\mid y_{i}>w_{n})=\mathbb{I}_{p+1}$
implies $\mathbb{E}(\|v_{i}\|^{2}\mid y_{i}>w_{n})=p+1$, so the Cauchy--Schwarz
inequality and the fourth-moment bound yield 
\[
\mathbb{E}\left(\left.\|v_{i}\|^{3}\right|y_{i}>w_{n}\right)\leq\left[\mathbb{E}\left(\left.\|v_{i}\|^{2}\right|y_{i}>w_{n}\right)\right]^{1/2}\left[\mathbb{E}\left(\left.\|v_{i}\|^{4}\right|y_{i}>w_{n}\right)\right]^{1/2}\leq C\underline{\lambda}_{n}^{-1}.
\]
Hence, using $\mathbb{E}(\epsilon_{ni}+1\mid y_{i}>w_{n},x_{i})=1$
and Markov's inequality, 
\[
\sup_{\|u\|\leq K}\left|\mathcal{R}_{n}(u)\right|=O_{P}\!\left(\frac{1}{\sqrt{n\Pr(Y>w_{n})}\,\underline{\lambda}_{n}}\right)=o_{P}(1)
\]
by (\ref{eq:varying_rank}). The conditional fourth-moment requirement
is used here to control the Taylor remainder after normalization by
$\bar{\varSigma}_{w_{n}}^{-1/2}$; note that the fourth-moment bound
alone would only give $\mathbb{E}(\|v_{i}\|^{3}\mid y_{i}>w_{n})\leq C\underline{\lambda}_{n}^{-3/2}$,
which would not suffice under (\ref{eq:varying_rank}).

We show below that 
\begin{equation}
\left[n\Pr(Y>w_{n})\right]^{-1/2}\bar{\varSigma}_{w_{n}}^{-1/2}\dot{\mathcal{K}}_{n}^{*}(0)\overset{d}{\longrightarrow}N(0,\mathbb{I}_{p+1}),\label{eq:Kn.}
\end{equation}
and 
\begin{equation}
\bar{\varSigma}_{w_{n}}^{-1/2}\left[\left[n\Pr(Y>w_{n})\right]^{-1}\ddot{\mathcal{K}}_{n}^{*}(0)\right]\bar{\varSigma}_{w_{n}}^{-1/2}\overset{P}{\longrightarrow}\mathbb{I}_{p+1}.\label{eq:Kn..->E}
\end{equation}
By the usual convexity argument applied to (\ref{eq:Taylor_expansion}),
(\ref{eq:Kn.}) and (\ref{eq:Kn..->E}) imply 
\[
\sqrt{n\Pr(Y>w_{n})}\,\bar{\varSigma}_{w_{n}}^{1/2}(\hat{\theta}-\theta^{*})=O_{P}(1).
\]
Since $\|\bar{\varSigma}_{w_{n}}^{-1/2}\|=\underline{\lambda}_{n}^{-1/2}$,
this also gives the slowest-direction bound 
\[
\|\hat{\theta}-\theta^{*}\|=O_{P}\!\left(\frac{1}{\sqrt{n\Pr(Y>w_{n})\underline{\lambda}_{n}}}\right).
\]
Note that this is only a bound determined by the weakest eigenvalue;
directions with larger eigenvalues converge faster.

We next use the first-order condition. A first-order expansion of
$\dot{\mathcal{K}}_{n}^{*}(\hat{\gamma})=0$ around zero gives 
\begin{align*}
0= & \;\left[n\Pr(Y>w_{n})\right]^{-1/2}\bar{\varSigma}_{w_{n}}^{-1/2}\dot{\mathcal{K}}_{n}^{*}(0)\\
 & +\left\{ \bar{\varSigma}_{w_{n}}^{-1/2}\left[\left[n\Pr(Y>w_{n})\right]^{-1}\ddot{\mathcal{K}}_{n}^{*}(0)\right]\bar{\varSigma}_{w_{n}}^{-1/2}\right\} \sqrt{n\Pr(Y>w_{n})}\,\bar{\varSigma}_{w_{n}}^{1/2}(\hat{\theta}-\theta^{*})+o_{P}(1).
\end{align*}
Therefore, (\ref{eq:Kn.}) and (\ref{eq:Kn..->E}) imply 
\begin{equation}
\sqrt{n\Pr(Y>w_{n})}\,\bar{\varSigma}_{w_{n}}^{1/2}(\hat{\theta}-\theta^{*})\overset{d}{\longrightarrow}N(0,\mathbb{I}_{p+1}).\label{eq:AN-population-Sigma}
\end{equation}

It remains to establish (\ref{eq:Kn.}) and (\ref{eq:Kn..->E}). Recall
the definition 
\begin{equation}
\epsilon_{ni}=\exp\left(x_{i}'\theta^{*}\right)\log\left(\left.y_{i}\right/w_{n}\right)-1.\label{eq:epsilon_ni}
\end{equation}
Conditional on $\{y_{i}>w_{n},x_{i}\}$, $\epsilon_{ni}+1$ is standard
exponential. Hence $\mathbb{E}(\epsilon_{ni}\mid y_{i}>w_{n},x_{i})=0$
and $\mathbb{E}(\epsilon_{ni}^{2}\mid y_{i}>w_{n},x_{i})=1$. Let
\[
q_{ni}=\left[n\Pr(Y>w_{n})\right]^{-1/2}\bar{\varSigma}_{w_{n}}^{-1/2}x_{i}\epsilon_{ni}I(y_{i}>w_{n}).
\]
Then $\mathbb{E}(q_{ni})=0$ and 
\begin{align}
\sum_{i=1}^{n}\mathbb{E}(q_{ni}q_{ni}') & =\frac{n\Pr(y_{i}>w_{n})}{n\Pr(Y>w_{n})}\bar{\varSigma}_{w_{n}}^{-1/2}\mathbb{E}(x_{i}x_{i}'\mid y_{i}>w_{n})\bar{\varSigma}_{w_{n}}^{-1/2}\nonumber \\
 & =\mathbb{I}_{p+1}.\label{eq:EZZ'}
\end{align}
For any fixed $(p+1)\times1$ vector $c$ with $\|c\|=1$, the moment
condition implies 
\begin{align}
\sum_{i=1}^{n}\mathbb{E}|c'q_{ni}|^{2+\delta} & \leq C\frac{\underline{\lambda}_{n}^{-(1+\delta/2)}}{\left[n\Pr(Y>w_{n})\right]^{\delta/2}}\mathbb{E}\left(\left.\|x_{i}\|^{2+\delta}\right|y_{i}>w_{n}\right)\nonumber \\
 & =o(1).\label{eq:x^2+delta}
\end{align}
To see the last equality, note that the intercept implies $\underline{\lambda}_{n}\leq1$,
and, for $\delta\geq2$, 
\[
\frac{\underline{\lambda}_{n}^{-(1+\delta/2)}}{\left[n\Pr(Y>w_{n})\right]^{\delta/2}}=\frac{\underline{\lambda}_{n}^{\delta/2-1}}{\left[n\Pr(Y>w_{n})\underline{\lambda}_{n}^{2}\right]^{\delta/2}}\leq\frac{1}{\left[n\Pr(Y>w_{n})\underline{\lambda}_{n}^{2}\right]^{\delta/2}}\to0
\]
by (\ref{eq:varying_rank}). Thus the Lyapunov condition holds in
every fixed direction. By the Cramér--Wold device, 
\[
\sum_{i=1}^{n}q_{ni}=\left[n\Pr(Y>w_{n})\right]^{-1/2}\bar{\varSigma}_{w_{n}}^{-1/2}\dot{\mathcal{K}}_{n}^{*}(0)\overset{d}{\longrightarrow}N(0,\mathbb{I}_{p+1}),
\]
which proves (\ref{eq:Kn.}).

For the Hessian, 
\begin{align}
\mathbb{E}\left[\left[n\Pr(Y>w_{n})\right]^{-1}\ddot{\mathcal{K}}_{n}^{*}(0)\right] & =\mathbb{E}\left[\left[n\Pr(Y>w_{n})\right]^{-1}\sum_{i=1}^{n}x_{i}x_{i}'(\epsilon_{ni}+1)I(y_{i}>w_{n})\right]\nonumber \\
 & =\bar{\varSigma}_{w_{n}}.\label{eq:Kn..}
\end{align}
The same fourth-moment calculation as in the original proof gives,
for each fixed pair $(j,l)$, 
\[
\operatorname{Var}\left[\left[n\Pr(Y>w_{n})\right]^{-1}\sum_{i=1}^{n}x_{ij}x_{il}(\epsilon_{ni}+1)I(y_{i}>w_{n})\right]=O(\left[n\Pr(Y>w_{n})\right]^{-1}).
\]
Since $p$ is fixed, 
\[
\left\Vert \left[n\Pr(Y>w_{n})\right]^{-1}\ddot{\mathcal{K}}_{n}^{*}(0)-\bar{\varSigma}_{w_{n}}\right\Vert =O_{P}(\left[n\Pr(Y>w_{n})\right]^{-1/2}).
\]
Consequently, 
\begin{align*}
 & \left\Vert \bar{\varSigma}_{w_{n}}^{-1/2}\left\{ \left[n\Pr(Y>w_{n})\right]^{-1}\ddot{\mathcal{K}}_{n}^{*}(0)-\bar{\varSigma}_{w_{n}}\right\} \bar{\varSigma}_{w_{n}}^{-1/2}\right\Vert \\
 & \qquad\leq\underline{\lambda}_{n}^{-1}O_{P}(\left[n\Pr(Y>w_{n})\right]^{-1/2})=O_{P}\!\left(\frac{1}{\sqrt{n\Pr(Y>w_{n})}\,\underline{\lambda}_{n}}\right)=o_{P}(1),
\end{align*}
which proves (\ref{eq:Kn..->E}).

Finally, define 
\[
\widetilde{\varSigma}_{w_{n}}=\left[n\Pr(Y>w_{n})\right]^{-1}\sum_{i=1}^{n}x_{i}x_{i}'I(y_{i}>w_{n}).
\]
The same fourth-moment calculation as for the Hessian gives 
\[
\left\Vert \widetilde{\varSigma}_{w_{n}}-\bar{\varSigma}_{w_{n}}\right\Vert =O_{P}\!\left(\left[n\Pr(Y>w_{n})\right]^{-1/2}\right).
\]
Moreover, $n_{0}$ is a sum of $n$ i.i.d. Bernoulli variables with
success probability $\Pr(Y>w_{n})$, so $\textrm{Var}\left(n_{0}\right)\leq n\Pr(Y>w_{n})$
and Chebyshev's inequality gives 
\[
\frac{n_{0}}{n\Pr(Y>w_{n})}-1=O_{P}\!\left(\left[n\Pr(Y>w_{n})\right]^{-1/2}\right).
\]
Since $\hat{\varSigma}_{w_{n}}=\frac{n\Pr(Y>w_{n})}{n_{0}}\widetilde{\varSigma}_{w_{n}}$,
and since $\|\bar{\varSigma}_{w_{n}}\|=\rho_{\max}(\bar{\varSigma}_{w_{n}})$
is uniformly bounded under the moment condition, the last two displays
yield 
\begin{equation}
\left\Vert \hat{\varSigma}_{w_{n}}-\bar{\varSigma}_{w_{n}}\right\Vert =O_{P}\!\left(\left[n\Pr(Y>w_{n})\right]^{-1/2}\right).\label{eq:Sigma-hat-rate}
\end{equation}
In particular, since $\|\bar{\varSigma}_{w_{n}}^{-1/2}\|=\underline{\lambda}_{n}^{-1/2}$,
\begin{equation}
\left\Vert \bar{\varSigma}_{w_{n}}^{-1/2}(\hat{\varSigma}_{w_{n}}-\bar{\varSigma}_{w_{n}})\bar{\varSigma}_{w_{n}}^{-1/2}\right\Vert \leq\underline{\lambda}_{n}^{-1}\left\Vert \hat{\varSigma}_{w_{n}}-\bar{\varSigma}_{w_{n}}\right\Vert =O_{P}\!\left(\frac{1}{\sqrt{n\Pr(Y>w_{n})}\,\underline{\lambda}_{n}}\right)=o_{P}(1),\label{eq:relative-Sigma-consistency}
\end{equation}
by (\ref{eq:varying_rank}).

It remains to replace the population normalization $\bar{\varSigma}_{w_{n}}^{1/2}$
in (\ref{eq:AN-population-Sigma}) by its sample counterpart $\hat{\varSigma}_{w_{n}}^{1/2}$.
We note that (\ref{eq:relative-Sigma-consistency}) by itself is not
sufficient for this step, because the matrix square root is not a
Lipschitz function of its argument and the condition number of $\bar{\varSigma}_{w_{n}}$
diverges when $\underline{\lambda}_{n}\to0$. We therefore use the
rate in (\ref{eq:Sigma-hat-rate}) together with Lemma \ref{LE:3}.
Since $\hat{\varSigma}_{w_{n}}$ is positive semidefinite by construction,
$\rho_{\min}(\hat{\varSigma}_{w_{n}})\geq0$, and Lemma \ref{LE:3}
applied with $A=\hat{\varSigma}_{w_{n}}$ and $B=\bar{\varSigma}_{w_{n}}$
gives 
\[
\left\Vert \hat{\varSigma}_{w_{n}}^{1/2}-\bar{\varSigma}_{w_{n}}^{1/2}\right\Vert _{F}\leq\frac{\left\Vert \hat{\varSigma}_{w_{n}}-\bar{\varSigma}_{w_{n}}\right\Vert _{F}}{\underline{\lambda}_{n}^{1/2}}.
\]
Because $p$ is fixed, the spectral and Frobenius norms are equivalent,
$\left\Vert \cdot\right\Vert \leq\left\Vert \cdot\right\Vert _{F}\leq\sqrt{p+1}\left\Vert \cdot\right\Vert $.
Multiplying on the right by $\bar{\varSigma}_{w_{n}}^{-1/2}$ and
using $\|\bar{\varSigma}_{w_{n}}^{-1/2}\|=\underline{\lambda}_{n}^{-1/2}$
therefore gives 
\begin{align}
\left\Vert \hat{\varSigma}_{w_{n}}^{1/2}\bar{\varSigma}_{w_{n}}^{-1/2}-\mathbb{I}_{p+1}\right\Vert  & =\left\Vert \left(\hat{\varSigma}_{w_{n}}^{1/2}-\bar{\varSigma}_{w_{n}}^{1/2}\right)\bar{\varSigma}_{w_{n}}^{-1/2}\right\Vert \leq C\frac{\left\Vert \hat{\varSigma}_{w_{n}}-\bar{\varSigma}_{w_{n}}\right\Vert }{\underline{\lambda}_{n}}\nonumber \\
 & =O_{P}\!\left(\frac{1}{\sqrt{n\Pr(Y>w_{n})}\,\underline{\lambda}_{n}}\right)=o_{P}(1),\label{eq:sqrt-ratio}
\end{align}
where the last line uses (\ref{eq:Sigma-hat-rate}) and (\ref{eq:varying_rank}).
Note that (\ref{eq:sqrt-ratio}) requires no condition beyond (\ref{eq:varying_rank}).

Finally, write the exact identity 
\[
\sqrt{n_{0}}\,\hat{\varSigma}_{w_{n}}^{1/2}(\hat{\theta}-\theta^{*})=\left(\frac{n_{0}}{n\Pr(Y>w_{n})}\right)^{1/2}\left(\hat{\varSigma}_{w_{n}}^{1/2}\bar{\varSigma}_{w_{n}}^{-1/2}\right)\sqrt{n\Pr(Y>w_{n})}\,\bar{\varSigma}_{w_{n}}^{1/2}(\hat{\theta}-\theta^{*}).
\]
The first factor converges in probability to one, the second converges
in probability to $\mathbb{I}_{p+1}$ by (\ref{eq:sqrt-ratio}), and
the third converges in distribution to $N(0,\mathbb{I}_{p+1})$ by
(\ref{eq:AN-population-Sigma}). Slutsky's theorem therefore yields
\[
\sqrt{n_{0}}\,\hat{\varSigma}_{w_{n}}^{1/2}(\hat{\theta}-\theta^{*})\overset{d}{\longrightarrow}N(0,\mathbb{I}_{p+1}),
\]
as desired. \hfill{}$\blacksquare$

\noindent\textbf{Proof of Theorem \ref{TH:density_quantile}.} (\ref{eq:quantile_equivalent})
and (\ref{eq:equivalent_condition}) imply that

\[
Q_{Y}\left(\tau|X_{1},X_{2}\right)=X_{1}+X_{2}\cdot Q_{\tilde{U}}\left(\tau|X_{1},X_{2}\right).
\]
Note that $Y=X_{1}+X_{2}\cdot\tilde{U}>w\Leftrightarrow\tilde{U}>\left.\left(w-X_{1}\right)\right/X_{2}.$
Using the above and $\bar{F}_{\tilde{U}}\left(\tilde{u}|X_{1},X_{2}\right)=\tilde{u}^{-\alpha},$
for $w>\bar{u}_{x_{1}}+\bar{u}_{x_{2}}$, 
\begin{align}
f\left(x_{1},x_{2}|Y>w\right) & =\frac{\Pr\left(Y>w\left|\left(X_{1},X_{2}\right)=\left(x_{1},x_{2}\right)\right.\right)f\left(x_{1},x_{2}\right)}{\Pr\left(Y>w\right)}\nonumber \\
 & =\frac{\left[\left.\left(w-x_{1}\right)\right/x_{2}\right]^{-\alpha}f\left(x_{1},x_{2}\right)}{\int_{\underline{u}_{x_{2}}}^{\bar{u}_{x_{2}}}\int_{\underline{u}_{x_{1}}}^{\bar{u}_{x_{1}}}\left[\left.\left(w-s_{1}\right)\right/s_{2}\right]^{-\alpha}f\left(s_{1},s_{2}\right)\textrm{d}s_{1}\textrm{d}s_{2}}.\label{eq:proof-tilt}
\end{align}
Write 
\[
\left[\left.\left(w-x_{1}\right)\right/x_{2}\right]^{-\alpha}=w^{-\alpha}x_{2}^{\alpha}\left(1-\frac{x_{1}}{w}\right)^{-\alpha},
\]
and multiply the numerator and the denominator of (\ref{eq:proof-tilt})
by $w^{\alpha}$, which gives 
\[
f\left(x_{1},x_{2}|Y>w\right)=\frac{x_{2}^{\alpha}\left(1-x_{1}/w\right)^{-\alpha}f\left(x_{1},x_{2}\right)}{\int_{\underline{u}_{x_{2}}}^{\bar{u}_{x_{2}}}\int_{\underline{u}_{x_{1}}}^{\bar{u}_{x_{1}}}s_{2}^{\alpha}\left(1-s_{1}/w\right)^{-\alpha}f\left(s_{1},s_{2}\right)\textrm{d}s_{1}\textrm{d}s_{2}}.
\]
Since the support of $X_{1}$ is bounded, 
\[
\sup_{x_{1}\in\left[\underline{u}_{x_{1}},\bar{u}_{x_{1}}\right]}\left|\left(1-\frac{x_{1}}{w}\right)^{-\alpha}-1\right|\rightarrow0\textrm{ as }w\rightarrow\infty,
\]
so the factor $\left(1-x_{1}/w\right)^{-\alpha}$ can be replaced
by one in both the numerator and the denominator, uniformly in $\left(x_{1},x_{2}\right)$.
Moreover, because $0<\underline{u}_{x_{2}}\leq x_{2}\leq\bar{u}_{x_{2}}<\infty$
and $f$ is a density that is bounded and bounded away from zero on
its support, 
\[
0<\underline{u}_{x_{2}}^{\alpha}\leq\int_{\underline{u}_{x_{2}}}^{\bar{u}_{x_{2}}}\int_{\underline{u}_{x_{1}}}^{\bar{u}_{x_{1}}}s_{2}^{\alpha}f\left(s_{1},s_{2}\right)\textrm{d}s_{1}\textrm{d}s_{2}\leq\bar{u}_{x_{2}}^{\alpha}<\infty,
\]
so the denominator is bounded away from zero and infinity. This proves
(\ref{eq:conditional_den_x1x2|w}). Note that only the compactness
of the support and $\underline{u}_{x_{2}}>0$ are used for the limit
itself; the lower bound on $f$ is used to conclude that the limiting
density is bounded away from zero.

For the uniform case, $f\left(x_{1},x_{2}\right)=\left[\left(\bar{u}_{x_{1}}-\underline{u}_{x_{1}}\right)\left(\bar{u}_{x_{2}}-\underline{u}_{x_{2}}\right)\right]^{-1}$,
so 
\[
\int_{\underline{u}_{x_{2}}}^{\bar{u}_{x_{2}}}\int_{\underline{u}_{x_{1}}}^{\bar{u}_{x_{1}}}s_{2}^{\alpha}f\left(s_{1},s_{2}\right)\textrm{d}s_{1}\textrm{d}s_{2}=\frac{\bar{u}_{x_{2}}^{\alpha+1}-\underline{u}_{x_{2}}^{\alpha+1}}{\left(\alpha+1\right)\left(\bar{u}_{x_{2}}-\underline{u}_{x_{2}}\right)}.
\]
Substituting this and $f\left(x_{1},x_{2}\right)$ into (\ref{eq:conditional_den_x1x2|w})
gives (\ref{eq:conditional_den_uniform}). \hfill{}$\blacksquare$

\section{On the Nonparametric Component in Tail Index Regressions}

\label{APP:B}

In this appendix, we study the effect of the degeneracy of $f(x\mid Y>w)$
on nonparametric tail index regressions when estimation is based on
exceedance samples $\{Y>w\}$ uniformly for all $x$. Focusing on
a degree-0 B-spline (histogram) approximation, we derive a pointwise
convergence-rate bound for the estimator. Note that this estimator
is equivalent to a kernel estimator with a uniform kernel, as in de
Haan and Zhou (2021), but differs in how the effective sample is defined
through the threshold used for estimation.

Our analysis makes explicit how the effective sample size within each
local neighborhood depends on $f(x\mid Y>w)$, and shows that the
convergence-rate bound deteriorates as this conditional density becomes
small. These results give a concrete illustration of how degeneracy
affects nonparametric estimation, and add to the discussion in the
main text.

\subsection*{Model and Conditions}

Suppose the following conditions hold. 
\begin{enumerate}
\item For $y\ge w_{n}>0$, 
\begin{equation}
\Pr(Y>y\mid X=x)=y^{-\alpha(x)},\label{eq:condition1}
\end{equation}
where $x\in[0,1]$, $\alpha(x)=\exp\{g(x)\}$, and $g$ is continuously
differentiable on $[0,1]$. 
\item Partition $[0,1]$ into $m_{n}$ equal subintervals $\{A_{j}\}_{j=1}^{m_{n}}$,
where $m_{n}\to\infty$. Define 
\[
g_{n}(x)=\sum_{j=1}^{m_{n}}\beta_{j}B_{j}(x),\quad B_{j}(x)=I\{x\in A_{j}\}.
\]
\item The number of partitions $m_{n}$ satisfies 
\begin{equation}
\frac{n\Pr(Y>w_{n})}{m_{n}}\inf_{x}f(x\mid Y>w_{n})\to\infty,\label{eq:condition3}
\end{equation}
as $w_{n}\to\infty$. 
\end{enumerate}
To illustrate the restriction in (\ref{eq:condition3}), consider
the uniform example underlying Corollary \ref{Coro:1}. Let $X\sim\operatorname{Uniform}[0,1]$
and $\alpha(x)=1+x$, or equivalently $g(x)=\log(1+x)$. Then $Z=\alpha(X)=1+X$
is uniformly distributed on $[1,2]$. Using the conditional-density
calculation in the proof of Corollary \ref{Coro:1}, we obtain 
\[
f(x\mid Y>w_{n})=\frac{w_{n}^{-x}\log w_{n}}{1-w_{n}^{-1}},\qquad x\in[0,1].
\]
Hence, 
\[
\inf_{x\in[0,1]}f(x\mid Y>w_{n})=\frac{\log w_{n}}{w_{n}(1-w_{n}^{-1})}\propto\frac{\log w_{n}}{w_{n}}.
\]
Condition (\ref{eq:condition3}) therefore becomes 
\[
\frac{n\Pr(Y>w_{n})\log w_{n}}{m_{n}w_{n}}\to\infty.
\]
Moreover, 
\[
\Pr(Y>w_{n})=\int_{0}^{1}w_{n}^{-(1+x)}\,dx=\frac{w_{n}^{-1}(1-w_{n}^{-1})}{\log w_{n}}.
\]
The same condition can equivalently be written as 
\[
\frac{n}{m_{n}w_{n}^{2}}\to\infty.
\]
Thus, (\ref{eq:condition3}) imposes a joint restriction on the growth
of $m_{n}$ and $w_{n}$ and can indeed become demanding as the conditional
density degenerates.

\subsubsection*{Degree-0 B-spline Estimator}

Let $\hat{g}_{n}(x)$ denote the degree-0 B-spline estimator obtained
by minimizing the approximate negative log-likelihood 
\[
Q_{n}(\beta)=\sum_{i=1}^{n}\Bigg[\exp\!\Big(\sum_{j=1}^{m_{n}}\beta_{j}B_{j}(x_{i})\Big)\log\Big(\frac{y_{i}}{w_{n}}\Big)-\sum_{j=1}^{m_{n}}\beta_{j}B_{j}(x_{i})\Bigg]I\{y_{i}>w_{n}\}.
\]

Since $B_{j}(x_{i})=1$ if $x_{i}\in A_{j}$ and $0$ otherwise, the
objective function can be rewritten as 
\[
Q_{n}(\beta)=\sum_{j=1}^{m_{n}}\sum_{i:x_{i}\in A_{j}}\Big[e^{\beta_{j}}\log(y_{i}/w_{n})-\beta_{j}\Big]I\{y_{i}>w_{n}\}.
\]

Let $N_{nj}=\sum_{i=1}^{n}I\{x_{i}\in A_{j},\,y_{i}>w_{n}\}$. Then
the minimizer has the closed-form expression 
\begin{align}
e^{\hat{\beta}_{j}} & =\left[\frac{1}{N_{nj}}\sum_{i:x_{i}\in A_{j},\,y_{i}>w_{n}}\log(y_{i}/w_{n})\right]^{-1}\nonumber \\
 & =\Bigg\{\mathbb{E}\!\left[\log(Y/w_{n})\mid Y>w_{n},\,X\in A_{j}\right]\nonumber \\
 & \quad+\frac{1}{N_{nj}}\sum_{i:x_{i}\in A_{j},\,y_{i}>w_{n}}\left(\log(y_{i}/w_{n})-\mathbb{E}\!\left[\log(Y/w_{n})\mid Y>w_{n},\,X\in A_{j}\right]\right)\Bigg\}^{-1}.\label{eq:e^beta}
\end{align}

\subsubsection*{Convergence Rate}

We analyze the two components in (\ref{eq:e^beta}).

For the first term, under (\ref{eq:condition1}) and a standard calculation
for Pareto-type tails, we obtain 
\begin{equation}
\mathbb{E}\left[\log(Y/w_{n})\mid Y>w_{n},X\in A_{j}\right]=e^{-g(x)}+O(m_{n}^{-1}),\quad\text{for any }x\in A_{j}.\label{eq:bias_term_b_spline}
\end{equation}

For the second term, by the i.i.d. assumption and standard moment
arguments (e.g., the Markov inequality), 
\begin{equation}
\frac{1}{N_{nj}}\sum_{i:x_{i}\in A_{j},\,y_{i}>w_{n}}\left(\log(y_{i}/w_{n})-\mathbb{E}\left[\log(Y/w_{n})\mid Y>w_{n},X\in A_{j}\right]\right)=O_{P}(N_{nj}^{-1/2}).\label{eq:variance_term_b_spline}
\end{equation}

Therefore, the key step is to find the rate of $N_{nj}$. Note that
\begin{align*}
\mathbb{E}(N_{nj}) & =\sum_{i=1}^{n}\Pr(Y_{i}>w_{n},X_{i}\in A_{j})\\
 & =n\Pr(Y>w_{n})\Pr(X\in A_{j}\mid Y>w_{n})\\
 & =n\Pr(Y>w_{n})\int_{A_{j}}f(x\mid Y>w_{n})\,dx.
\end{align*}
Hence, 
\[
\frac{n\Pr(Y>w_{n})}{m_{n}}\inf_{x\in A_{j}}f(x\mid Y>w_{n})\leq\mathbb{E}(N_{nj})\leq\frac{n\Pr(Y>w_{n})}{m_{n}}\sup_{x\in A_{j}}f(x\mid Y>w_{n}).
\]

Under the rate condition (\ref{eq:condition3}), arguments similar
to Theorem 1 of Wang and Tsai (2009) yield 
\[
\frac{N_{nj}}{\mathbb{E}(N_{nj})}=1+o_{P}(1),\qquad\frac{n_{0}}{n\Pr(Y>w_{n})}=1+o_{P}(1).
\]

Combining the above results, we obtain 
\begin{equation}
\frac{n_{0}}{m_{n}}\inf_{x\in A_{j}}f(x\mid Y>w_{n})\,[1+o_{P}(1)]\leq N_{nj}\leq\frac{n_{0}}{m_{n}}\sup_{x\in A_{j}}f(x\mid Y>w_{n})\,[1+o_{P}(1)].\label{eq:N_b_0_spline}
\end{equation}

Since $e^{\hat{\beta}_{j}}$ estimates $e^{g(x)}$ for $x\in A_{j}$,
we write $\hat{g}(x)=\hat{\beta}_{j}$ for $x\in A_{j}$. Substituting
(\ref{eq:bias_term_b_spline}) and (\ref{eq:variance_term_b_spline})
into (\ref{eq:e^beta}) gives 
\[
e^{\hat{g}(x)}-e^{g(x)}=O_{P}\left(N_{nj}^{-1/2}+m_{n}^{-1}\right).
\]

Applying (\ref{eq:N_b_0_spline}) then yields 
\begin{equation}
\hat{g}(x)-g(x)=O_{P}\left(\sqrt{\frac{m_{n}}{n_{0}}}\left[\inf_{x\in A_{j}}f(x\mid Y>w_{n})\right]^{-1/2}+m_{n}^{-1}\right),\quad x\in A_{j}.\label{eq:rate_degree_0}
\end{equation}

Thus, the pointwise convergence-rate bound in (\ref{eq:rate_degree_0})
deteriorates as the conditional density $f(x\mid Y>w_{n})$ becomes
small in the neighborhood containing $x$.

\section{Estimation of $\mathbb{E}\left(XX'|Y>w\right)$ under Weak Dependence}

\label{APP:C}

\citet{CarrascoChen2002} established the $\beta$-mixing property
and the existence of higher-order moments for a range of GARCH models
under regularity conditions. In this appendix, we establish consistency
of the sample Gram matrix under $\alpha$-mixing, which is weaker
than $\beta$-mixing, together with suitable uniform moment and rate
conditions. We do not require the observations to be identically distributed.

For possibly non-identically distributed observations, define 
\[
p_{i,n}=\Pr(Y_{i}>w_{n}),\qquad\bar{p}_{n}=\frac{1}{n}\sum_{i=1}^{n}p_{i,n},
\]
and 
\[
\bar{\varSigma}_{w_{n}}=\frac{\sum_{i=1}^{n}\mathbb{E}\left[X_{i}X_{i}'I(Y_{i}>w_{n})\right]}{\sum_{i=1}^{n}\Pr(Y_{i}>w_{n})}.
\]
Thus, $\bar{\varSigma}_{w_{n}}$ is the natural population counterpart
of 
\[
\hat{\varSigma}_{w_{n}}=\frac{\sum_{i=1}^{n}X_{i}X_{i}'I\left(Y_{i}>w_{n}\right)}{\sum_{i=1}^{n}I\left(Y_{i}>w_{n}\right)}.
\]
When the observations are identically distributed, $\bar{\varSigma}_{w_{n}}=\mathbb{E}(XX'\mid Y>w_{n})$,
so the above reduces to the notation used in the main text. Assume
that $X_{i}$ is $(p+1)\times1$, with fixed dimension $p$, and let
$X_{il}$ denote its $l$-th component. We impose the following conditions. 
\begin{enumerate}
\item The sequence $\{(X_{i},Y_{i})\}_{i\in\mathbb{Z}}$ is $\alpha$-mixing,
not necessarily stationary or identically distributed. Define the
uniform mixing coefficients 
\[
\varrho(h)=\sup_{k\in\mathbb{Z}}\alpha\!\left(\sigma\{(X_{i},Y_{i}):i\leq k\},\sigma\{(X_{i},Y_{i}):i\geq k+h\}\right).
\]
For some $\delta>0$, suppose 
\begin{equation}
\sum_{h=1}^{\infty}\varrho(h)^{\delta/(1+\delta)}<\infty.\label{eq:alpha_mixing_coef}
\end{equation}
\item The conditional moments are uniformly bounded across calendar time:
\[
\sup_{n\geq1}\max_{1\leq i\leq n}\max_{1\leq l\leq p+1}\mathbb{E}\left(\left.|X_{il}|^{4+4\delta}\right|Y_{i}>w_{n}\right)<\infty.
\]
\item Let 
\[
\underline{\lambda}_{n}=\rho_{\min}(\bar{\varSigma}_{w_{n}}).
\]
As $w_{n}\to\infty$, suppose 
\begin{equation}
n\bar{p}_{n}^{(1+2\delta)/(1+\delta)}\underline{\lambda}_{n}^{2}\longrightarrow\infty.\label{eq:weak_dependence_rate}
\end{equation}
No common rate is imposed on the remaining eigenvalues. 
\end{enumerate}
To illustrate the restriction in (\ref{eq:weak_dependence_rate}),
consider again the uniform example of Corollary \ref{Coro:1}, now
viewed as a weakly dependent array: $\alpha(X_{i})=Z_{i}$ is uniformly
distributed on $[\underline{u},\bar{u}]$ with $\underline{u}>0$,
so that $\underline{\lambda}_{n}\propto(\log w_{n})^{-2}$ and $\Pr(Y_{i}>w_{n})\propto1/[(w_{n})^{\underline{u}}\log w_{n}]$,
and hence $\bar{p}_{n}\propto(w_{n})^{-\underline{u}}/\log w_{n}$.
Writing 
\[
\kappa=\frac{1+2\delta}{1+\delta}\in(1,2),
\]
condition (\ref{eq:weak_dependence_rate}) becomes 
\begin{equation}
\frac{(\log w_{n})^{\kappa+4}(w_{n})^{\kappa\underline{u}}}{n}\to0.\label{eq:weak_dependence_example}
\end{equation}
Setting $\kappa=1$ gives exactly the independent-sampling condition
obtained for the same example in the remark on condition (\ref{eq:varying_rank})
in Section \ref{sec:2}. The dependence enters only through the exponent
$\kappa>1$, which increases both the polynomial factor $(w_{n})^{\underline{u}}$
and the logarithmic factor.

We claim that for any $\varepsilon>0$, 
\begin{equation}
\Pr\left\{ \sup_{\|c\|=1}\left|\frac{c'\hat{\varSigma}_{w_{n}}c}{c'\bar{\varSigma}_{w_{n}}c}-1\right|>\varepsilon\right\} \longrightarrow0,\qquad n\to\infty.\label{eq:uniform_converge_X'X}
\end{equation}

We show the claim below. For the proof, $C$ denotes a generic positive
constant that may vary from line to line. Define 
\[
\hat{p}_{n}=\frac{1}{n}\sum_{i=1}^{n}I(Y_{i}>w_{n}),\qquad\hat{M}_{n}=\frac{1}{n}\sum_{i=1}^{n}X_{i}X_{i}'I(Y_{i}>w_{n}),
\]
and their expectations 
\[
\mathbb{E}(\hat{p}_{n})=\bar{p}_{n},\qquad M_{n}=\mathbb{E}(\hat{M}_{n})=\bar{p}_{n}\bar{\varSigma}_{w_{n}}.
\]
Then $\hat{\varSigma}_{w_{n}}=\hat{M}_{n}/\hat{p}_{n}$.

We first establish a rate for the denominator. Since $\operatorname{Var}[I(Y_{i}>w_{n})]\leq p_{i,n}$,
the diagonal terms satisfy 
\[
\frac{1}{n^{2}}\sum_{i=1}^{n}\operatorname{Var}[I(Y_{i}>w_{n})]\leq\frac{\bar{p}_{n}}{n}.
\]
For $i<j$, Davydov's inequality with exponents $2+2\delta,2+2\delta,$
and $(1+\delta)/\delta$ gives 
\[
\left|\operatorname{Cov}\left(I(Y_{i}>w_{n}),I(Y_{j}>w_{n})\right)\right|\leq C\varrho(j-i)^{\delta/(1+\delta)}(p_{i,n}p_{j,n})^{1/(2+2\delta)}.
\]
For a fixed lag $h$, the Cauchy--Schwarz inequality and Jensen's
inequality imply 
\begin{align*}
\sum_{i=1}^{n-h}(p_{i,n}p_{i+h,n})^{1/(2+2\delta)} & \leq\left(\sum_{i=1}^{n}p_{i,n}^{1/(1+\delta)}\right)^{1/2}\left(\sum_{i=1}^{n}p_{i,n}^{1/(1+\delta)}\right)^{1/2}\\
 & \leq n\bar{p}_{n}^{1/(1+\delta)}.
\end{align*}
Hence, using (\ref{eq:alpha_mixing_coef}), 
\[
\operatorname{Var}(\hat{p}_{n})\leq\frac{C}{n}\bar{p}_{n}^{1/(1+\delta)}.
\]
Therefore, 
\begin{equation}
\frac{\hat{p}_{n}}{\bar{p}_{n}}-1=O_{P}\left(\frac{1}{\sqrt{n\bar{p}_{n}^{(1+2\delta)/(1+\delta)}}}\right).\label{eq:denominator_rate}
\end{equation}

We next establish the same rate for the numerator. Consider its $(l,m)$-th
element and write 
\[
Z_{i,lm}=X_{il}X_{im}I(Y_{i}>w_{n}).
\]
By the uniform conditional moment condition and the Cauchy--Schwarz
inequality, 
\begin{align*}
\mathbb{E}|Z_{i,lm}|^{2+2\delta} & =p_{i,n}\,\mathbb{E}\left(\left.|X_{il}X_{im}|^{2+2\delta}\right|Y_{i}>w_{n}\right)\\
 & \leq Cp_{i,n}.
\end{align*}
Thus, another application of Davydov's inequality yields 
\[
|\operatorname{Cov}(Z_{i,lm},Z_{j,lm})|\leq C\varrho(j-i)^{\delta/(1+\delta)}(p_{i,n}p_{j,n})^{1/(2+2\delta)}.
\]
The same summation argument used for the denominator therefore gives
\[
\operatorname{Var}\left(\frac{1}{n}\sum_{i=1}^{n}Z_{i,lm}\right)\leq\frac{C}{n}\bar{p}_{n}^{1/(1+\delta)}.
\]
Since $p$ is fixed, element-wise convergence implies 
\begin{equation}
\|\hat{M}_{n}-M_{n}\|=O_{P}\left(\sqrt{\frac{\bar{p}_{n}^{1/(1+\delta)}}{n}}\right).\label{eq:numerator_rate}
\end{equation}

The uniform moment condition also implies 
\[
\sup_{n}\|\bar{\varSigma}_{w_{n}}\|<\infty,
\]
because 
\[
\operatorname{tr}(\bar{\varSigma}_{w_{n}})=\frac{\sum_{i=1}^{n}p_{i,n}\,\mathbb{E}(\|X_{i}\|^{2}\mid Y_{i}>w_{n})}{\sum_{i=1}^{n}p_{i,n}}\leq C.
\]
Condition (\ref{eq:weak_dependence_rate}) therefore implies $n\bar{p}_{n}^{(1+2\delta)/(1+\delta)}\to\infty$,
and hence $\hat{p}_{n}/\bar{p}_{n}\overset{P}{\to}1$ by (\ref{eq:denominator_rate}).
Combining (\ref{eq:denominator_rate}) and (\ref{eq:numerator_rate})
gives 
\begin{align}
\|\hat{\varSigma}_{w_{n}}-\bar{\varSigma}_{w_{n}}\| & =\left\Vert \frac{\hat{M}_{n}}{\hat{p}_{n}}-\frac{M_{n}}{\bar{p}_{n}}\right\Vert \nonumber \\
 & =O_{P}\left(\frac{1}{\sqrt{n\bar{p}_{n}^{(1+2\delta)/(1+\delta)}}}\right).\label{eq:gram_rate_weak_dependence}
\end{align}

Finally, 
\begin{align*}
\sup_{\|c\|=1}\left|\frac{c'\hat{\varSigma}_{w_{n}}c}{c'\bar{\varSigma}_{w_{n}}c}-1\right| & =\sup_{\|c\|=1}\left|\frac{c'(\hat{\varSigma}_{w_{n}}-\bar{\varSigma}_{w_{n}})c}{c'\bar{\varSigma}_{w_{n}}c}\right|\\
 & \leq\underline{\lambda}_{n}^{-1}\|\hat{\varSigma}_{w_{n}}-\bar{\varSigma}_{w_{n}}\|\\
 & =O_{P}\left(\frac{1}{\sqrt{n\bar{p}_{n}^{(1+2\delta)/(1+\delta)}}\,\underline{\lambda}_{n}}\right)=o_{P}(1),
\end{align*}
by (\ref{eq:weak_dependence_rate}). This proves (\ref{eq:uniform_converge_X'X}).

We end with a comment on the rate obtained above. Under independent
sampling, the argument in the proof of Theorem \ref{TH:2} gives $\|\hat{\varSigma}_{w_{n}}-\bar{\varSigma}_{w_{n}}\|=O_{P}\left(\left[n\bar{p}_{n}\right]^{-1/2}\right)$,
and the corresponding rate condition is $n\bar{p}_{n}\underline{\lambda}_{n}^{2}\to\infty$.
Condition (\ref{eq:weak_dependence_rate}) replaces $\bar{p}_{n}$
by $\bar{p}_{n}^{(1+2\delta)/(1+\delta)}$ and is therefore strictly
stronger, since $(1+2\delta)/(1+\delta)\in(1,2)$ and $\bar{p}_{n}\to0$.
The loss comes from the covariance bound rather than from the dependence
itself: Davydov's inequality bounds the covariance of two rare-event
terms by $\varrho(h)^{\delta/(1+\delta)}\left(p_{i,n}p_{j,n}\right)^{1/(2+2\delta)}$,
and the factor $\left(p_{i,n}p_{j,n}\right)^{1/(2+2\delta)}$ understates
how rare the event is, compared with the variance term, which is of
order $\bar{p}_{n}$.

 \bibliographystyle{ecta}
\bibliography{references}

\end{document}